\documentclass[sigconf]{acmart}

\copyrightyear{2026}
\acmYear{2026}
\setcopyright{cc}
\setcctype{by}
\acmConference[CCS '26]{Proceedings of the 2026 ACM SIGSAC Conference on Computer and Communications Security}{November 15--19, 2026}{The Hague, Netherlands}
\acmBooktitle{Proceedings of the 2026 ACM SIGSAC Conference on Computer and Communications Security (CCS '26), November 15--19, 2026, The Hague, Netherlands}
\acmDOI{10.1145/3830454.3832621}
\acmISBN{979-8-4007-2871-6/2026/11}

\usepackage{amsmath,amsopn,amsthm}
\usepackage{subfigure}
\usepackage{xspace,fancyvrb,multirow}
\usepackage[T1]{fontenc}

\usepackage{url}
\usepackage{fancyhdr}
\usepackage{enumitem}
\usepackage{fp}
\usepackage{siunitx}
\usepackage{balance}
\usepackage{listings}
\usepackage{soul}
\usepackage{fontawesome}
\usepackage{bbding, pifont}

\usepackage{amsmath}
\usepackage{colortbl}      
\usepackage{makecell}
\usepackage{algorithm}
\usepackage{algorithmic}
\usepackage{xspace}   
\usepackage{stfloats}
\usepackage{fvextra}

\newcommand{\eg}{\textit{e.g., }}
\newcommand{\ie}{\textit{i.e., }}

\newcommand{\rust}{Rust\xspace}
\newcommand{\unsafe}{unsafe\xspace}
\newcommand{\srust}{safe Rust\xspace}
\newcommand{\usrust}{unsafe Rust\xspace}
\newcommand{\unknownbugs}{\textsc{13}\xspace}

\newcommand{\cc}[1]{\texttt{#1}}

\newcommand{\Fref}[1]{Figure~\hyperref[#1]{\ref{#1}}}
\newcommand{\Tref}[1]{Table~\hyperref[#1]{\ref{#1}}}
\newcommand{\Sref}[1]{\hyperref[#1]{\textsection\ref{#1}}}
\newcommand{\Aref}[1]{Algorithm~\hyperref[#1]{\ref{#1}}}
\newcommand{\APref}[1]{Appendix~\hyperref[#1]{\textsection\ref{#1}}}

\newcommand{\sys}{\textsc{RustGo}\xspace}
\newcommand{\rustgobase}{\textsc{RG-base}\xspace}
\newcommand{\rustgoopt}{\textsc{RG-opt}\xspace}
\newcommand{\rustgoprun}{\textsc{RG-prun}\xspace}

\newcommand{\rustgodissafe}{\textsc{DisSafe}\xspace}
\newcommand{\rustgosync}{\textsc{RG-Sync}\xspace}

\newcommand{\benchnum}{13\xspace}

\newcommand{\optimizedtargetrate}{\textsc{84.13\%}\xspace}

\newcommand{\totalbb}{\textsc{40,252}\xspace}
\newcommand{\stdbb}{\textsc{24,924}\xspace}
\newcommand{\nonstdbb}{\textsc{15,329}\xspace}

\newcommand{\rustgobasebb}{\textsc{19,605}\xspace}
\newcommand{\rustgooptbb}{\textsc{18,895}\xspace}

\newcommand{\rustgobb}{\textsc{8,743}\xspace}

\newcommand{\rustgobasepr}{\textsc{51.11\%}\xspace}
\newcommand{\rustgobaseprapproximate}{\textsc{27.38\%}\xspace} 
\newcommand{\rustgooptprapproximate}{\textsc{25.65\%}\xspace} 

\newcommand{\rustgooptpr}{\textsc{52.84\%}\xspace}

\newcommand{\rustgopr}{\textsc{78.49\%}\xspace}

\newcommand{\totalaveragespeedup}{$\times$ 3.33\xspace}

\newcommand{\afltime}{1,098,621\xspace}
\newcommand{\afl}{$\times$ 3.45\xspace}
\newcommand{\aflttr}{$\times$ 3.83\xspace}
\newcommand{\afldstime}{523,749\xspace}
\newcommand{\afldsrate}{$\times$ 5.29\xspace}
\newcommand{\aflpptime}{722,828\xspace}
\newcommand{\aflpp}{$\times$ 2.85\xspace}
\newcommand{\aflppttr}{$\times$ 3.24\xspace}
\newcommand{\aflppnewvul}{$\times$ 4.72\xspace}
\newcommand{\aflppnewvultime}{758,539\xspace}
\newcommand{\aflppdstime}{300,197\xspace}
\newcommand{\aflppdsrate}{$\times$ 3.49\xspace}
\newcommand{\aflgotime}{1,710,052\xspace}
\newcommand{\aflgo}{$\times$ 4.30\xspace}
\newcommand{\aflgottr}{$\times$ 5.50\xspace}
\newcommand{\aflgodstime}{466,484\xspace}
\newcommand{\aflgodsrate}{$\times$ 5.47\xspace}
\newcommand{\windrangertime}{1,830,377\xspace}
\newcommand{\windranger}{$\times$ 5.08\xspace}
\newcommand{\windrangerttr}{$\times$ 6.65\xspace}
\newcommand{\windrangerdstime}{409,838\xspace}
\newcommand{\windrangerdsrate}{$\times$ 6.56\xspace}
\newcommand{\fishfuzztime}{1,135,880\xspace}
\newcommand{\fishfuzz}{$\times$ 4.15\xspace}
\newcommand{\fishfuzzttr}{$\times$ 3.44\xspace}
\newcommand{\fishfuzzdstime}{507,384\xspace}
\newcommand{\fishfuzzdsrate}{$\times$ 5.35\xspace}
\newcommand{\aflruntime}{505,195\xspace}
\newcommand{\aflrun}{$\times$ 2.11\xspace}
\newcommand{\aflrunttr}{$\times$ 2.15\xspace}
\newcommand{\aflrunnewvul}{$\times$ 3.72\xspace}
\newcommand{\aflrunnewvultime}{832,945\xspace}
\newcommand{\aflrundstime}{290,500\xspace}
\newcommand{\aflrundsrate}{$\times$ 2.51\xspace}
\newcommand{\lysotime}{712,234\xspace}
\newcommand{\lyso}{$\times$ 2.61\xspace}
\newcommand{\lysottr}{$\times$ 2.30\xspace}
\newcommand{\lysodstime}{261,478\xspace}
\newcommand{\lysodsrate}{$\times$ 3.25\xspace}
\newcommand{\panickillertime}{598,031\xspace}
\newcommand{\panickiller}{$\times$ 2.09\xspace}
\newcommand{\panickillerttr}{$\times$ 2.04\xspace}
\newcommand{\panickillernewvul}{$\times$ 4.89\xspace}
\newcommand{\panickillernewvultime}{4,093,334\xspace}
\newcommand{\panickillerdstime}{278,546\xspace}
\newcommand{\panickillerdsrate}{$\times$ 3.09\xspace}
\newcommand{\rustgotime}{312,542\xspace}

\newcommand{\rustgonewvultime}{408,294\xspace}
\newcommand{\rustgodstime}{161,769\xspace}

\newcommand{\rustgodissafetime}{1,353,207\xspace}
\newcommand{\rustgodissaferesult}{$\times$ 3.27\xspace}
\newcommand{\rustgodissafettr}{$\times$ 4.10\xspace}
\newcommand{\rustgodissafenewvul}{$\times$ 4.71\xspace}
\newcommand{\rustgodissafenewvultime}{698,714\xspace}
\newcommand{\rustgodissafedstime}{231,232\xspace}
\newcommand{\rustgodissafedsrate}{$\times$ 3.87\xspace}
\newcommand{\rustgobasetime}{1,005,920\xspace}
\newcommand{\rustgobaseresult}{$\times$ 3.07\xspace}
\newcommand{\rustgobasettr}{$\times$ 3.60\xspace}
\newcommand{\rustgobasenewvul}{$\times$ 3.82\xspace}
\newcommand{\rustgobasenewvultime}{689,721\xspace}
\newcommand{\rustgobasedstime}{218,022\xspace}
\newcommand{\rustgobasedsrate}{$\times$ 3.59\xspace}
\newcommand{\rustgoopttime}{737,259\xspace}
\newcommand{\rustgooptresult}{$\times$ 2.60\xspace}
\newcommand{\rustgooptttr}{$\times$ 2.88\xspace}
\newcommand{\rustgooptnewvul}{$\times$ 3.40\xspace}
\newcommand{\rustgooptnewvultime}{651,804\xspace}
\newcommand{\rustgooptdstime}{201,118\xspace}
\newcommand{\rustgooptdsrate}{$\times$ 3.16\xspace}
\newcommand{\rustgopruntime}{424,543\xspace}
\newcommand{\rustgoprunresult}{$\times$ 1.36\xspace}
\newcommand{\rustgoprunttr}{$\times$ 1.29\xspace}
\newcommand{\rustgoprunnewvul}{$\times$ 3.59\xspace}
\newcommand{\rustgoprunnewvultime}{4,034,454\xspace}
\newcommand{\rustgoprundstime}{177,659\xspace}
\newcommand{\rustgoprundsrate}{$\times$ 2.02\xspace}
\newcommand{\rustgosynctime}{431,924\xspace}
\newcommand{\rustgosyncresult}{$\times$ 1.36\xspace}
\newcommand{\rustgosyncttr}{$\times$ 1.24\xspace}
\newcommand{\rustgosyncnewvul}{$\times$ 3.59\xspace}
\newcommand{\rustgosyncnewvultime}{4,034,454\xspace}
\newcommand{\rustgosyncdstime}{178,015\xspace}
\newcommand{\rustgosyncdsrate}{$\times$ 2.04\xspace}

\newcommand{\rustsecnum}{six\xspace}

\begin{document}

\title{\sys: Fairly Directed Greybox Fuzzing for Enforcing Rust Memory Safety}


\author{Dongyeon Yu}
\authornote{Equal contribution.}
\affiliation{%
  \institution{Korea University}
  \city{Seoul}
  \country{Republic of Korea}
}
\email{dy3199@korea.ac.kr}

\author{Jiun Min}
\authornotemark[1]
\affiliation{%
  \institution{Korea University}
  \city{Seoul}
  \country{Republic of Korea}
}
\email{jiunmin@korea.ac.kr}

\author{Yewan Na}
\affiliation{%
  \institution{Ulsan National Institute of Science and Technology}
  \city{Ulsan}
  \country{Republic of Korea}
}
\email{ytri3055@unist.ac.kr}

\author{Mijung Kim}
\affiliation{%
  \institution{Ulsan National Institute of Science and Technology}
  \city{Ulsan}
  \country{Republic of Korea}
}
\email{mijungk@unist.ac.kr}

\author{Taegyu Kim}
\affiliation{%
  \institution{The Pennsylvania State University}
  \city{University Park}
  \state{Pennsylvania}
  \country{USA}
}
\email{tgkim@psu.edu}

\author{Yuseok Jeon}
\authornote{Corresponding author.}
\affiliation{%
  \institution{Korea University}
  \city{Seoul}
  \country{Republic of Korea}
}
\email{ys\_jeon@korea.ac.kr}

\begin{abstract}
    \rust is a popular systems programming language that provides strong memory safety and introduces low-performance overhead.
While \rust guarantees memory safety through strict security policies, such as ownership, memory bugs can still occur in \unsafe-related \rust codes where these policies are not fully enforced.
Although such \usrust code accounts for only a small portion of the entire code (\eg 10\%), existing approaches fuzz the entire code---including \srust, whose memory safety is already enforced by the \rust compiler---resulting in inefficient use of fuzzing resources.
  
In this paper, we propose \sys, the new \rust-directed greybox fuzzer that effectively and fairly  
focuses on code regions potentially containing memory bugs.
For this, \sys automatically identifies potential memory bug targets and accurately prunes the paths irrelevant to each target by leveraging \rust-specific static analysis.
For each identified target, \sys includes a new fuzzing approach that maintains an independent state and applies dynamic pruning to maximize balanced and focused fuzzing.
We evaluate \sys on various 
real-world \rust applications. On average, \sys prunes \rustgopr of irrelevant paths, reaches targets \panickiller to \windranger faster than existing fuzzers, and identifies \unknownbugs unknown bugs (\rustsecnum assigned RUSTSEC IDs and one assigned CVE ID).

\end{abstract}


\renewcommand{\shortauthors}{Dongyeon Yu et al.}

\maketitle

\section{Introduction}
Unlike existing C/C++ systems programming languages, \rust ~\cite{matsakis2014rust,web:rust-instroduction} guarantees memory safety with low-performance overhead by strictly applying ownership~\cite{web:ownership}, borrowing~\cite{web:borrowcheck}, and lifetime~\cite{web:lifetime} rules at compile time, along with bound checks at both compile time and runtime.
Due to these strengths, the White House issued guidelines~\cite{web:whitehouse,web:whitehouse-memory-safe} in 2024 recommending the use of memory-safe languages, such as \rust, one of the most effective solutions against memory bugs. 
Additionally, the adoption of \rust continues to expand across major platforms~\cite{ RustLinux, web:RustChrome}. For example, Android now includes several major components~\cite{web:RustAndroid} implemented in \rust, such as DNS-over-HTTP3~\cite{web:RustDNSOVERHTTP} and the Android Virtualization Framework (AVF)~\cite{web:RustAndroidVirtualization}, with approximately 1.5 million lines~\cite{web:RustAndroidLine} of \rust code currently present in the Android Open Source Project (AOSP)~\cite{web:RustAndroidOpenSource}.
The Linux kernel, previously implemented only in C, also officially adopted \rust~\cite{web:LinuxKernelRust} as the second development language starting from version 6.1 in October 2022.

However, despite \rust's security benefits, \rust's strict memory safety policies restrict~\cite{jung2017rustbelt, wang2019towards} the implementation of certain operations, including low-level operation (\eg raw-pointer dereference) and the creation of complex data structures (\eg doubly-linked lists). 
To address these limitations, \rust introduces the unsafe regions~\cite{UnsafeRust} to bypass specific safety policies when necessary.
However, since Rust’s security policies do not fully apply to these unsafe code-related regions~\cite{UnsafeRust}, memory bugs can still occur within those regions. 
According to RustSec~\cite{web:rustsec}, Rust’s security vulnerability database, all reported memory bugs (22\% of the 581 vulnerabilities reported between 2016 and 2023) are caused by incorrect use of unsafe code.

In light of these security reports, although memory bugs are not widespread across all of \rust's codes, they still appear in unsafe-related regions. Thus, efficient and effective testing of these regions is necessary to identify and eliminate memory bugs.
For this, fuzzing~\cite{miller1990empirical}, generally combined with sanitizers (\eg Address Sanitizer~\cite{serebryany2012addresssanitizer}), is widely used to detect memory bugs.
The \rust community has developed \rust fuzzers, such as afl.rs~\cite{web:afl.rs}, cargo-fuzz~\cite{web:cargo-fuzz}, and hongfuzz-rs~\cite{web:hong-fuzz}, on top of those existing fuzzers. 

However, these existing \rust fuzzers are originally designed for C/C++, with the assumption that memory bugs can occur anywhere in the program. Additionally, the previous study~\cite{astrauskas2020programmers, zheng2023closer, xu2021memory} shows that only about 10\% of a typical \rust program is written in unsafe code, while the remaining 90\% is already memory-safe, thanks to \rust's memory safety policies.
This highlights the need for Directed Greybox Fuzzing (DGF)~\cite{bohme2017directed, huang2022beacon, li2023gfuzz, osterlund2020parmesan, shah2022mc2, zheng2023fishfuzz} that directs fuzzing to the code regions where memory bugs may exist, instead of wasting fuzzing resources on the regions that are mostly memory-safe.
However, most existing DGF solutions are not specifically designed for \rust. 
While existing DGFs (\eg distance-based and pruning-based DGFs)—originally developed for memory-unsafe languages, such as C/C++—can be adapted to \rust, there remains significant room for improvement by leveraging \rust's unique features. 
To optimize \rust program fuzzing, there are three major challenges.

The first challenge is to automatically and precisely identify and optimize targets in \rust programs where memory bugs are likely to occur.
However, most existing DGFs only rely on traditional criteria for target selection, such as code patches~\cite{bohme2013regression, marinescu2013katch}, new commits~\cite{xiang2024critical}, sanitizer check locations~\cite{zheng2023fishfuzz, osterlund2020parmesan}, or manual target selection~\cite{jin2012bugredux, pham2015hercules, christakis2016guiding, mathis2017detecting}.
Since there are no approaches capable of automatically identifying memory-bug-prone targets in \rust, we need new approaches with selection criteria.

Furthermore, a previous study~\cite{min2024erasan} shows that a typical \rust program contains, on average, approximately 10,197 unsafe memory access points that require memory safety checks.
Attempting to target all of these targets can lead to redundant path exploration and diminished fuzzing directionality.
As a result, DGFs may ultimately work as generic coverage-guided fuzzers without being properly directed to designated targets. 
Therefore, an automated and efficient approach to target selection and optimization is necessary for scalable and precise analysis.

The second challenge is path pruning.
Even when the target is precisely identified, minimizing the exploration of irrelevant paths remains difficult.
Unlike C/C++, \rust statically merges and links all codes by default, including the standard library, into a single binary.
This complicates precise analysis by blurring the boundary between user code and standard library code and increases the chance of over-approximation during reachability analysis, which is critical for effective path pruning.
For example, we observe that applying existing general-purpose reachability analysis to \rust leads to over-approximation: approximately 28\% of the reachable paths identified are, in fact, unrelated to the actual target.

The last challenge is ensuring balanced fuzzing efforts across all reachable targets.
Existing DGFs generally prioritize close targets, making them tested frequently, while distant targets are tested infrequently.
Moreover, shared fuzzing states (\eg a common input queue) make it harder to fairly allocate fuzzing energy across multiple targets.
Therefore, a new fuzzing framework is required to enable fair fuzzing across reachable targets.

To address these three challenges, we propose \sys, the new fuzzing framework tailored to \rust programs to detect \rust-specific memory bugs accurately and efficiently.
\sys (i) automatically identifies and optimizes targets where memory bugs may exist, (ii) performs target-specific path pruning by reflecting Rust’s unique characteristics, and (iii) pursues fairness by dynamically adjusting fuzzing focus and managing independent fuzzing states for each target.

More specifically, to address the first challenge---optimized target selection---\sys performs \rust-specific cross-IR analysis to identify all possible target points that may have memory bugs and then optimize them using inter-procedural taint and post-dominance analysis.
For the second challenge---precise irrelevant path elimination---\sys introduces target-specific path pruning analysis, which accurately prunes unnecessary code paths even in the presence of Rust’s complex standard library structure.
Lastly, to handle the last challenge of fair fuzzing energy allocation, \sys proposes a dynamic toggling approach that enables dynamic precise path pruning for only the currently selected targets, while managing independent fuzzing states for each target (\eg via multi-queue and per-target coverage bitmap), allowing for fair fuzzing opportunity across all reachable targets.

Our evaluation shows that \sys accurately identifies and merges targets and reaches them faster than existing state-of-the arts DGFs.
For evaluation, due to the lack of a fuzzing benchmark like MAGMA~\cite{hazimeh2020magma} for the \rust environments, we create a MAGMA-style benchmark by selecting \benchnum diverse applications via Principal Component Analysis (PCA)~\cite{web:pca}, similar to MAGMA.
Based on this benchmark, \sys identifies suspicious memory bug targets and merges \optimizedtargetrate of them through optimized target selection to avoid redundant testing. 
Additionally, \sys's target-specific pruning analysis prunes \rustgopr irrelevant code per target.
As a result, \sys detects (up to \windranger) vulnerabilities \windranger, \fishfuzz, \aflgo, \afl, \aflpp, \aflrun, \lyso, and \panickiller faster than WindRanger~\cite{du2022windranger}, FishFuzz~\cite{zheng2023fishfuzz}, AFLGO~\cite{bohme2017directed}, AFL~\cite{web:AFL}, AFL++~\cite{web:AFL++}, AFLRUN~\cite{rong2024toward}, Lyso~\cite{lyso}, and PanicKiller~\cite{panickiller}, respectively.
\sys also identifies \unknownbugs unknown memory bugs in real-world \rust projects, \rustsecnum of which have been assigned RUSTSEC IDs and one a CVE ID.

This paper makes the following contributions:
\begin{itemize}[itemsep=0em,topsep=0em,leftmargin=*]

\item
We propose the new \rust-specific directed greybox fuzzer that effectively and fairly focuses on code regions potentially containing memory bugs.

\item
\sys prunes \rustgopr of irrelevant paths and reaches targets \panickiller to \windranger faster than existing DGF fuzzers. 

\item
\sys identifies \unknownbugs previously unknown real-world memory bugs, of which ten have been confirmed, eight have been fixed, and \rustsecnum have been assigned RustSec IDs and one a CVE ID.

\end{itemize}

\section{Background and Motivation}
\subsection{Memory Bugs in \rust}
\label{ss:memorybugsinrust}
\rust guarantees memory safety by leveraging strict security rules, such as ownership~\cite{web:ownership}, lifetime inference~\cite{web:lifetime}, borrow checking~\cite{web:borrowcheck}, and bound checking~\cite{web:boundcheck}. 
However, Rust’s safety policies cannot be fully enforced in all situations.
For example, Rust's strict memory safety policies make it difficult to implement certain features, such as doubly linked lists or low-level memory operations like raw pointer dereferencing.
To address this, \rust introduces unsafe blocks~\cite{UnsafeRust}, which allow developers to bypass \rust's safety policies.
Therefore, memory safety is not strictly enforced within these regions, making them a common source of memory bugs.

According to the RustSec Advisory Database~\cite{web:rustsec}, approximately 22\% (131 of 581) of the security vulnerabilities reported between 2016 and 2023 are memory bugs~\cite{min2024erasan}.
These memory bugs commonly involve raw pointer dereferencing within \usrust.
This is because Rust's safety policies (\ie ownership, lifetimes, borrow, and bound checks) are enforced in most cases, although raw pointer dereferencing operates outside these guarantees.
In the case of spatial memory safety bugs, a raw pointer can hold an arbitrary address, and \rust does not apply compile-time or runtime bounds checks exclusively for raw pointer dereferencing.
Therefore, spatial memory safety bugs occur when a raw pointer accesses memory outside of its valid bounds.
Temporal memory safety bugs occur when freed memory is accessed through raw pointers or their aliased safe references.
More specifically, such bugs can occur when a raw pointer dereferences a freed object or when a safe reference (\eg \cc{\&} or \cc{\&mut}) accesses memory that has previously been freed through the raw pointer.

As a result, both spatial and temporal memory safety bugs originate from \usrust—more precisely, from regions related to raw pointer dereferencing.
Furthermore, since raw pointer dereferencing is disallowed in \srust, spatial memory safety bugs, which rely solely on raw pointer dereferencing, can only occur within \usrust.
In contrast, temporal memory safety bugs can occur even in \srust, as safe references may access objects that have been freed through raw pointers.

In real-world \rust applications, there exist a significant number of raw pointer dereferencing related code areas where memory bugs may appear.
The previous study~\cite{min2024erasan} reports that such locations amount to an average of 10,197 per program, based on an analysis of 23 widely used applications ranked by downloads and stars.
Given their large volume, manually locating all these potential memory bug areas is highly challenging.
Moreover, the abundance of such targets can lead to inefficient fuzzing, as many redundant paths may be explored.
Therefore, to enable efficient fuzzing, it is necessary to automatically and precisely identify the relevant targets and merge the targets to eliminate testing redundancy.

\subsection{\rust Fuzzers} 

Existing fuzzers for testing \rust are mainly categorized into coverage-based fuzzers and fuzz driver generators. 
Among these, afl.rs~\cite{web:afl.rs}, cargo-fuzz~\cite{web:cargo-fuzz}, honggfuzz-rs~\cite{web:hong-fuzz} are coverage-guided fuzzers aiming to maximize code coverage.
However, these approaches do not fully consider the \rust language-specific characteristics particularly that memory bugs arise in unsafe-related codes leading to inefficient fuzzing by unnecessarily testing memory-safe codes. 
PanicKiller~\cite{panickiller} prioritizes fuzzing energy toward unsafe code regions with fewer runtime safety checks and mutates inputs to bypass assertions, thereby enabling the exploration of paths beyond panic assertions.
However, PanicKiller's static identification of numerous unsafe regions (\eg 10,197 per program)---without accounting for \rust's standard library---leads to target over-approximation, weakening its directionality.

The \rust fuzz driver generators~\cite{xu2024rpg,jiang2021rulf,syrust,crabtree,fries,rumono,rug,deepsurf} automatically generate corresponding fuzz drivers to test \rust library functions.
Among these, Syrust~\cite{syrust}, Crabtree~\cite{crabtree}, Fries~\cite{fries}, RULF~\cite{jiang2021rulf}, and Rumono~\cite{rumono} generate fuzz drivers for APIs defined in \rust libraries without considering unsafe code regions.
RPG~\cite{xu2024rpg} identifies library functions that use unsafe code and automatically generates corresponding fuzz drivers. However, since RPG targets functions that involve unsafe code, it may not detect memory bugs, such as use-after-free bugs, that can occur in safe Rust code.
Furthermore, RPG does not provide a fuzzing framework for fairly and efficiently identifying and fuzzing the numerous memory bug-prone locations within target applications.
RUG~\cite{rug} and deepSURF~\cite{deepsurf} adopt LLM-based techniques to generate compilable fuzz drivers and to handle complex Rust specific types (\eg generics and traits).
However, these approaches are limited in generating fuzz drivers that can detect all memory violation vulnerabilities, as they fail to adequately consider complicated vulnerability-triggering execution context (\eg UAF bugs) and handle complex argument types (\eg \cc{HashTable}) in vulnerable API in real-world applications.
Consequently, \rust needs a directed greybox fuzzing framework that can precisely identify and fairly focus fuzzing efforts on these memory bug-prone codes.

\subsection{Directed Greybox Fuzzing} 
\label{ss:dgfbg}
Directed Greybox Fuzzing (DGF)~\cite{bohme2017directed, huang2022beacon, xiang2024critical} is designed to prioritize fuzzing efforts toward selected targets (\eg patch validation) rather than covering as much code as possible.
These existing DGF approaches can be broadly categorized into three types: (i) distance-based DGF~\cite{bohme2017directed, chen2018hawkeye, lee2021constraint, shah2022mc2, du2022windranger}, (ii) pruning-based DGF~\cite{huang2022beacon, srivastava2022one, zong2020fuzzguard, luo2023selectfuzz, kim2023dafl}, and (iii) multi-target-supporting DGF~\cite{xiang2024critical, osterlund2020parmesan, huang2024titan, zheng2023fishfuzz, rong2024toward, lyso}.

\noindent \textbf{Distance-based DGF.}
Distance-based DGFs\cite{bohme2017directed, chen2018hawkeye, lee2021constraint, shah2022mc2, du2022windranger} improve the directness of the fuzzing by prioritizing intermediate program locations that are closer to the target.
For example, AFLGo~\cite{bohme2017directed} defines a distance metric and calculates the distance between a test input and the target to guide the fuzzer to select inputs that are more likely to reach targets.
Follow-up approaches, including Hawkeye~\cite{chen2018hawkeye}, which leverages call trace similarity, and WindRanger~\cite{du2022windranger}, which considers deviation basic blocks, further improve the accuracy of distance estimation.
While effective for single-target scenarios, existing distance-based DGF approaches tend to concentrate fuzzing energy on nearby targets when multiple target are considered, due to the use of a harmonic mean to aggregate distances across all targets.
Therefore, current DGF approaches do not ensure equitable fuzzing coverage across multiple targets when the number of targets increases.

\noindent \textbf{Pruning-based DGF.} Pruning-based DGF approaches~\cite{huang2022beacon, srivastava2022one, zong2020fuzzguard, luo2023selectfuzz, kim2023dafl} enhance fuzzing efficiency by eliminating paths and inputs that cannot likely reach the target.
For example, SieveFuzz~\cite{srivastava2022one} dynamically analyzes reachability during execution and discards inputs that exclusively reach unrelated code regions.
Beacon~\cite{huang2022beacon} performs reachability analysis and extracts path constraints, inserting pruning assertions to terminate fuzzing early when these assertion conditions are violated.
FuzzGuard~\cite{zong2020fuzzguard} leverages deep learning to predict the likelihood of reaching the target and prioritizes inputs accordingly.
SelectFuzz~\cite{luo2023selectfuzz} and DAFL~\cite{kim2023dafl} apply selective feedback by collecting coverage only from target-relevant code blocks, thereby implicitly suppressing the exploration of unrelated code.
However, since these pruning-based approaches are originally developed for C/C++, directly applying them to \rust results in over-approximated pruning.
More specifically, the complex structure of \rust including its all statically linked standard library and rich abstractions can lead to over-approximated analysis, causing many unreachable paths to remain unpruned.
Furthermore, when handling multiple targets, performing reachability and path constraint analysis across all targets collectively undermines pruning accuracy.
Additionally, there are currently no pruning approaches that consider these specific characteristics of memory bugs in \rust. 

\noindent \textbf{Multi-target-supporting DGF.}
Multi-target DGFs~\cite{xiang2024critical, osterlund2020parmesan, huang2024titan, zheng2023fishfuzz, rong2024toward, lyso}, unlike single-target DGFs, are designed to support fuzzing multiple targets simultaneously.
ParmeSan~\cite{osterlund2020parmesan} computes the harmonic mean of distances to all targets, prioritizing inputs that are relatively close to all of them.
WAFLGo~\cite{xiang2024critical} utilizes separate distance metrics per target and selects the closest.
Titan~\cite{huang2024titan} analyzes the path conditions and correlations among targets to select inputs likely to cover multiple targets at once.
FishFuzz~\cite{zheng2023fishfuzz} prioritizes common locations that lead to multiple targets, and AFLRun~\cite{rong2024toward} focuses its fuzzing energy on internally defined critical blocks.
%
Lyso~\cite{lyso} adopts a multi-target, multi-step guidance strategy that prioritizes alarms using semantic execution steps derived from static analysis.
%
However, current multi-target DGFs, which prioritize targets based on predefined polices, tend to focus fuzzing efforts on a limited set of targets, leaving others with little or no testing opportunity.
Therefore, a new DGF framework is required to support fair fuzzing, ensuring that all potentially vulnerable codes are thoroughly tested.
\section{Design}

\begin{figure*}[t]
    \centering
    \includegraphics[width=0.8\linewidth]{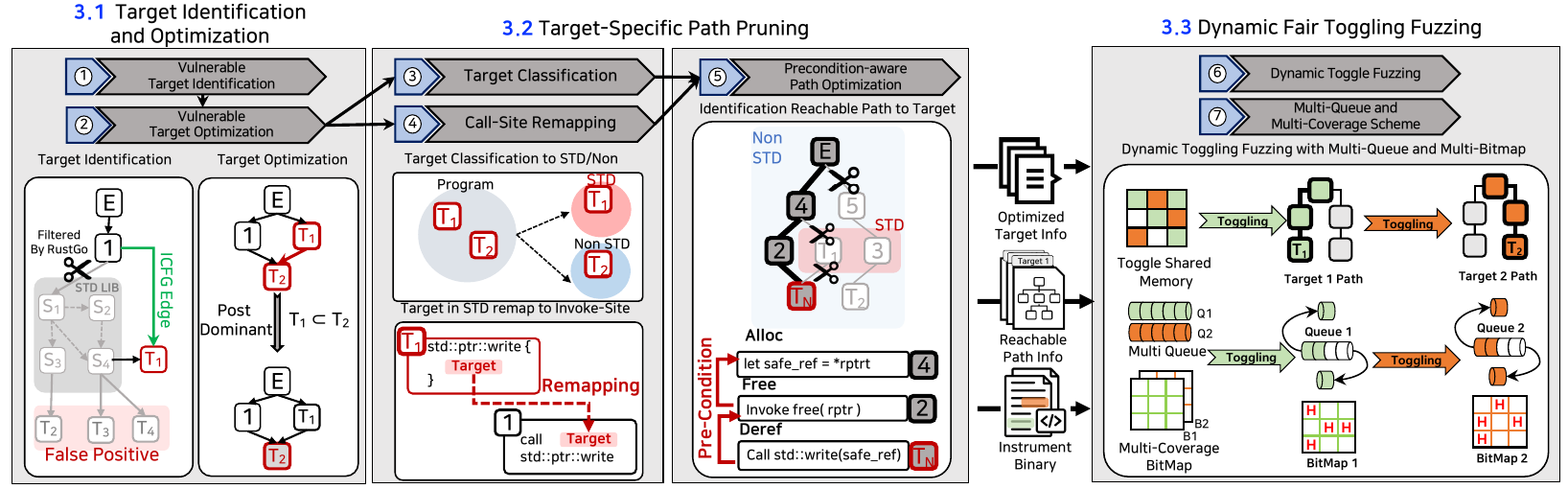}
    \caption{Overview of \sys.}
    \label{fig:3-overview}
\end{figure*}

We present the design of \sys, the new directed fuzzer for efficient memory bug detection within \rust applications. For that, \sys automatically identifies vulnerable targets and effectively distributes fuzzing energy across all such vulnerable targets. 
\Fref{fig:3-overview} illustrates the overall architecture of \sys. 
\sys consists of the three main components as follows.
\begin{enumerate}[label=(\roman*)]
    \item \textbf{Target Identification and Optimization} (\Sref{ss:DesignModule1}): \sys automatically identifies vulnerable targets and optimizes those targets for efficient path pruning.
    
    \item \textbf{Target-Specific Path Pruning} (\Sref{ss:DesignModule2}): For each identified target, \sys performs the target-specific reachability analysis tailored to a \rust application to statically eliminate paths that cannot trigger bugs or unreachable to each target.
    
    \item \textbf{Dynamic Fair Toggling Fuzzing} (\Sref{ss:DesignModule3}): \sys applies the dynamic toggling approach to balance the fuzzing energy across all targets.

\end{enumerate}

\subsection{Target Identification and Optimization}
\label{ss:DesignModule1}
Identifying vulnerable targets in \rust programs is challenging, as simply treating all \unsafe code blocks as fuzzing targets is overly coarse-grained, since memory bugs are not confined to \unsafe code. As discussed in \Sref{ss:memorybugsinrust}, raw pointers and aliased references can bypass \rust's safety guarantees and cause memory errors even in \srust (\eg RUSTSEC-2020-0097~\cite{RUSTSEC-2020-0097}).
Thus, simply targeting unsafe blocks can cause unnecessary checking of unsafe regions that do not actually contain bugs and overlook bugs that can occur in safe code; therefore, this section presents \sys's precise target identification approach.

In this paper, we define vulnerable targets as code location where \rust's memory safety guarantees can be bypassed, potentially leading to invalid memory access. For spatial bug, this refers to raw pointer dereferences to invalid addresses; for temporal bug, it refers to accesses to deallocated objects via raw pointers or aliased safe references. \sys identifies such targets through MIR-level type analysis, LLVM IR-based alias analysis, and VFG-based taint tracking. For spatial bugs, it detects raw pointer dereferences via MIR type matching. For temporal bugs, \sys performs alias analysis on raw pointers and uses VFG-based taint analysis to trace their propagation, enabling the detection of dereferences that may occur after deallocation.

Additionally, even when targets are precisely identified, a large number of selected targets can significantly increase testing costs and time. To efficiently test all bug-prone targets without false negatives, \sys employs Rust-specific target optimization approaches that leverage \rust's lifetime model and post-dominant target consolidation.

\subsubsection{Vulnerable Target Identification}
\label{ss:FPProblem}
To identify vulnerable target locations, \sys employs both MIR~\cite{web:rust-mir} and LLVM IR-level analyses.
These two levels of analysis are needed because raw pointers are fundamentally involved in memory safety bugs in \rust, yet this raw pointer information is explicitly available only at MIR and is not preserved after lowering to LLVM IR.
Thus, at the MIR level, \sys performs MIR-level type-matching analysis~\cite{min2024erasan} to identify operations involving raw pointers and annotates the corresponding instructions (\eg \cc{!rawptr}) to preserve this information across analysis stages.

At the LLVM IR level, \sys performs Value Flow Graph (VFG)~\cite{sui2016svf}-based taint analysis starting from annotated raw pointers to track their value-flow propagation and identify alias pointer dereference sites, \ie other potential targets where bugs can occur. 
However, because the above target identification is performed on the unified LLVM IR generated by the \rust compiler, it inevitably analyzes both \rust standard library code and non-standard library code together. Without explicitly handling the \rust standard library, this can introduce false positive issues, selecting targets that cannot actually trigger bugs.

These issues primarily arise because commonly used points-to analyses are context-insensitive, as context-sensitive variants are not scalable~\cite{cho2024rustsan, min2024erasan, bang2023trust}.
More specifically, as standard library functions (\eg \cc{drop}) appear pervasively in the unified LLVM IR, context-insensitive points-to analysis may incorrectly connect unrelated call sites through shared standard library functions, including code paths that are unreachable in actual execution.
This imprecision is further amplified by the deep nesting and extensive internal calls within the \rust standard library.
While inlining could theoretically mitigate these context-sensitivity issues, we find it ineffective in practice. More specifically, as shown in ~\Tref{tbl:drop-backup-data} in the Appendix \Sref{apen:evaluation-result-details}, we compared the IR before and after LLVM's inlining pass and observed that only 3.05\% of standard library functions are actually inlined; in particular, 99.68\% of drop-related functions remain as call instructions rather than being inlined.

\begin{figure}[t]
    \centering
    \includegraphics[width=0.5\textwidth]{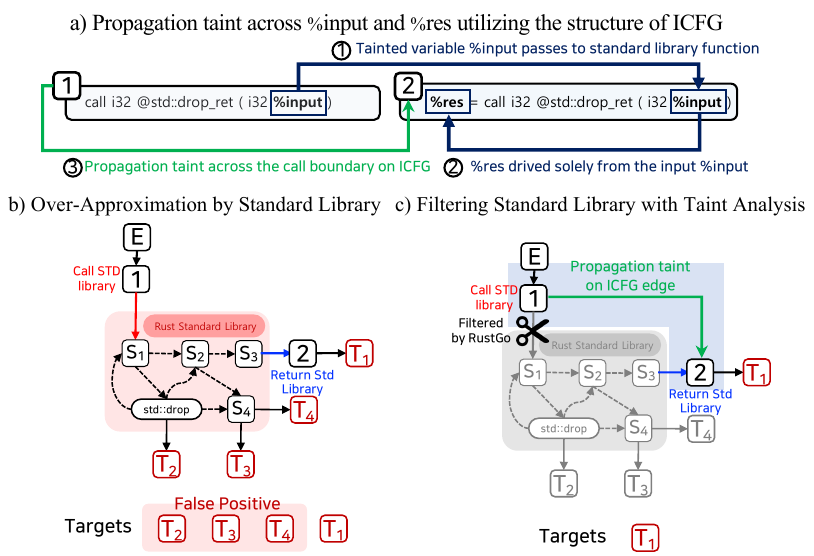}
    \caption{Overview of \sys's inter-procedural taint propagation across standard library calls. Red edges represent \cc{Call} edges and blue edges denote \cc{Return} edges. Green edges represent abstract taint propagation across standard library boundaries without traversing internal library logic.}
    \label{fig:taint-analysis}
\end{figure}

Notably, this problem persists even when \rust applications are compiled with dynamic linking (\eg \cc{-C prefer-dynamic}). Although dynamic linking aims to separate standard library code, most of the \rust standard library is generic and is   monomorphized\allowbreak~\cite{monomorphization} into the application's LLVM IR at compile time, leading to unavoidable over-approximation.
Summarization-based context-sensitive pointer analysis~\cite{wilson1995efficient} is also less suitable for \rust environment as generic code and standard library functions are monomorphized into type-specific instances (\eg \cc{drop<Vec<i32>\,>} or \cc{drop\allowbreak<String>)}. This significantly increases the number of function instances, making summarization-based approaches less scalable. For example, across our 13 benchmark programs, the average number of \cc{drop} instances is 540 in ~\Tref{tbl:drop-backup-data} in the Appendix \Sref{apen:evaluation-result-details}, which leads to a large number of summaries that need to be maintained and applied.

Therefore, to address these false positive issues, \sys introduces a false-positive target elimination approach that combines standard library filtering with inter-procedural taint analysis. 
As a first step, \sys performs standard library annotation by automatically referencing the definitions of \rust standard library functions (available in \cc{rust/library/std}). Through this process, \sys can automatically distinguish and annotate standard library functions from non-standard functions.

After annotating standard library functions, \sys performs inter-procedural taint analysis to identify vulnerable targets. To avoid false negatives (\eg removing targets that are actually reachable in real execution), \sys conservatively handles indirect calls and uninstantiated generics that may obscure call relationships or executable paths.
For example, \sys handles dynamic dispatch via trait objects by conservatively resolving vtable-based calls and linking each call site to all possible implementations identified from the LLVM IR, ensuring that no valid execution targets are pruned, even though this may result in some degree of over-approximation. 
\sys is also designed to naturally support generics based on \rust's monomorphization model. Specifically, in \rust, only the type instantiations actually used in the program are materialized in the IR (covered by \sys). Since \rust does not create new generic instances at runtime, uninstantiated types do not correspond to executable paths and therefore do not introduce false negatives.

Additionally, \sys applies customized inter-procedural taint analysis to effectively identify vulnerable targets.
Specifically, when tainted variables are passed as arguments to standard library functions, \sys intentionally avoids propagating taint through their internal VFG to prevent the analysis from exploring infeasible paths inside the standard library that are not reachable from the actual call context.
Instead, upon encountering a standard library call, our analysis bypasses the internal VFG and directly resumes propagation to the caller by referring to the Inter-procedural Control Flow Graph (ICFG).
More specifically, by leveraging ICFG edges that connect the call site to the return site, \sys maps taint between arguments and return values, thereby eliminating only unreachable paths without introducing false negatives.
This prevention enables \sys to filter out false-positive targets and improve the scalability of the analysis by reducing unnecessary taint propagation.

For example, as illustrated in \Fref{fig:taint-analysis}-(a), when a tainted variable (\eg \cc{\%input}) is passed to a standard library function (\eg \cc{\%res = call i32 @std::drop\_ret (i32 \%input)}), \sys avoids traversing the internal logic of standard library functions, which would otherwise include targets that are not actually reachable (\eg \Fref{fig:taint-analysis}-(b)).
For this, as shown in \Fref{fig:taint-analysis}-(c), \sys utilizes the structure of the ICFG to propagate taint across the call boundary: since the call site provides \cc{\%input} as an argument and receives \cc{\%res} as the return value, and the ICFG explicitly links the call site to the function's return site, \sys maps the taint from \cc{\%input} to \cc{\%res} by following this ICFG edge.

It is important to note that this standard library filtering does not mean that \sys ignores standard library code. \sys analyzes the full program IR, including the \rust standard library code available at compilation time, and identifies raw-pointer-related operations within both safe and unsafe regions as fuzzing targets. Thus, all memory-bug-prone codes related to the standard library remain within \sys's target-identification scope, including complex cases such as \cc{Vec::get()} and \cc{Vec::push()}, which involve raw pointer usage across safe/unsafe \rust interactions. In summary, this standard library abstraction in \sys is applied only during static analysis to prevent unnecessary exploration of infeasible paths through the standard library.

\begin{algorithm}[t]
\caption{Post-Dominant Target Consolidation in \sys}
\label{algo:algorithm1}

\begin{algorithmic}[1]

\STATE \textbf{Input:}
Set of target locations $T = \{t_1,t_2,\ldots,t_n\}$

\STATE \textbf{Output:}
Consolidated set of targets $T'$

\STATE \textbf{function} \textsc{ConsolidateTargets}$(T)$

\STATE
\hspace{\algorithmicindent}
$T' \gets \emptyset$

\STATE
\hspace{\algorithmicindent}
$\mathit{DomMap} \gets \{\}$

\STATE
\hspace{\algorithmicindent}
\textbf{for all} $(t_i,t_j) \in T \times T$
where $i \neq j$ \textbf{do}

\STATE
\hspace{\algorithmicindent}
\hspace{\algorithmicindent}
$R_i \gets \operatorname{ReachableSet}(t_i)$

\STATE
\hspace{\algorithmicindent}
\hspace{\algorithmicindent}
$R_j \gets \operatorname{ReachableSet}(t_j)$

\STATE
\hspace{\algorithmicindent}
\hspace{\algorithmicindent}
\textbf{if} $R_i \subset R_j$ \textbf{then}

\STATE
\hspace{\algorithmicindent}
\hspace{\algorithmicindent}
\hspace{\algorithmicindent}
$\mathit{DomMap}[t_i] \gets t_j$

\STATE
\hspace{\algorithmicindent}
\hspace{\algorithmicindent}
\textbf{else if} $R_j \subset R_i$ \textbf{then}

\STATE
\hspace{\algorithmicindent}
\hspace{\algorithmicindent}
\hspace{\algorithmicindent}
$\mathit{DomMap}[t_j] \gets t_i$

\STATE
\hspace{\algorithmicindent}
\hspace{\algorithmicindent}
\textbf{end if}

\STATE
\hspace{\algorithmicindent}
\textbf{end for}

\STATE
\hspace{\algorithmicindent}
$\mathit{Groups}
 \gets \operatorname{MergeByDominator}(\mathit{DomMap})$

\STATE
\hspace{\algorithmicindent}
\textbf{for all} group
$G \in \mathit{Groups}$ \textbf{do}

\STATE
\hspace{\algorithmicindent}
\hspace{\algorithmicindent}
$t_r \gets \operatorname{SelectRepresentative}(G)$

\STATE
\hspace{\algorithmicindent}
\hspace{\algorithmicindent}
$T' \gets T' \cup \{t_r\}$

\STATE
\hspace{\algorithmicindent}
\textbf{end for}

\STATE
\hspace{\algorithmicindent}
\textbf{return} $T'$

\STATE \textbf{end function}

\end{algorithmic}
\end{algorithm}

\begin{figure}[t]
    \centering
    \includegraphics[width=0.45\textwidth]{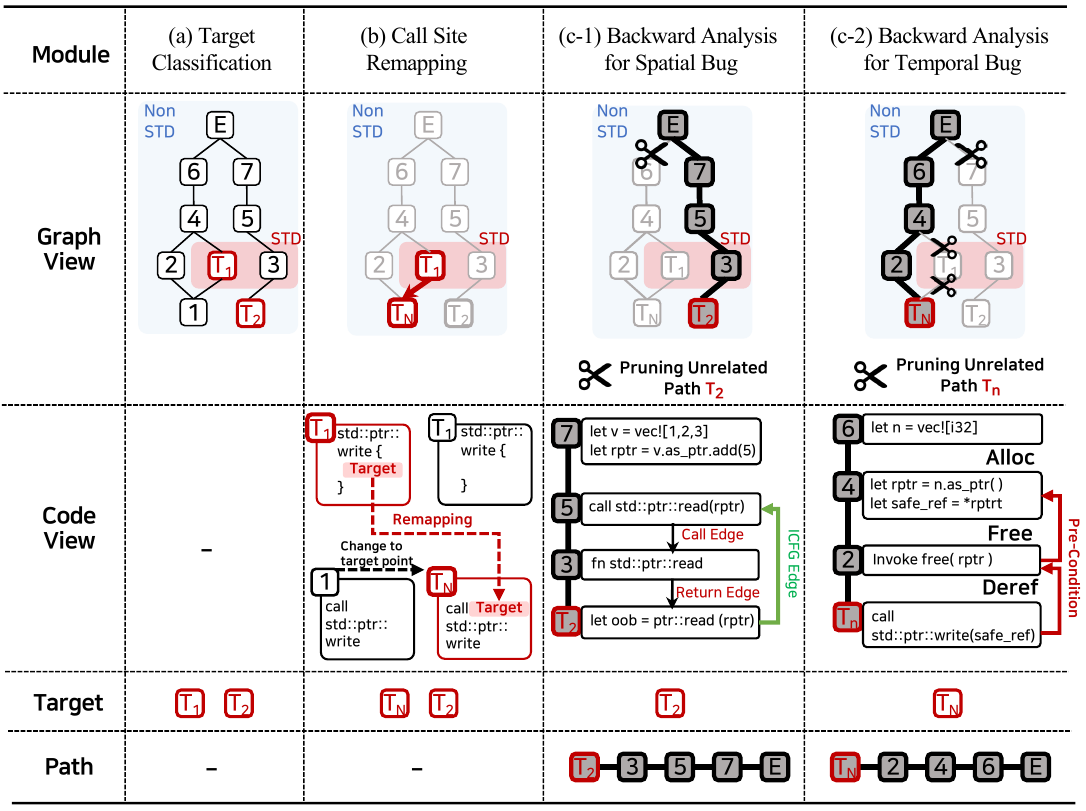}
    \caption{Overview of target-specific path pruning analysis. Each box represents the basic block. Basic blocks located in the blue region correspond to code from non-standard libraries, while those in the red region originate from \rust standard libraries. }
    \label{fig:path-pruning-analysis}
\end{figure}

\subsubsection{Vulnerable Target Optimization}
Despite the accurate vulnerable target identification in the prior step, the number of the identified targets can still remain large. 
Such many targets may cause DGFs to fail to direct targets during fuzzing, making DGFs work almost similar to coverage-guided fuzzing or even worse, as discussed in \Sref{ss:dgfbg}.
This eventually leads to fuzzing performance degradation~\cite{xiang2024critical, he2023rltg, zheng2023fishfuzz, rong2024toward, liang2023multiple}.

To mitigate this, \sys applies post-dominance analysis to merge redundant targets. 
Specifically, \sys merges target points when a post-dominance relationship exists between them.
Following the notion of post-dominators~\cite{gupta1992generalized}, a target \( t_i \) is post-dominated by another target \( t_j \) if all paths from \( t_i \) are also reachable from \( t_j \). 
For example, in such cases, covering \( t_j \) implicitly ensures the coverage of \( t_i \), allowing \( t_i \) and \( t_j \) to be merged into a single effective target.
While post-dominance is a general control-flow concept, existing post-dominance analyse implementations (\eg LLVM~\cite{web:llvm-14.0.6} and SVF~\cite{sui2016svf}) are currently limited to the intra-procedural level. 
As a result, post-dominance relationships are computed only within individual functions, causing many opportunities to merge targets to be missed. To address this limitation, \sys introduces an inter-procedural basic-block post-dominance analysis. Instead of relying on traditional post-dominator trees, \sys infers post-dominance relationships by comparing inter-procedural backward reachable basic-block sets for each target.
This enables \sys to apply post-dominance reasoning across function boundaries and effectively merge redundant targets.

\Aref{algo:algorithm1} shows our post-dominance analysis to consolidate redundant fuzzing targets. \sys first performs backward reachability analysis (\Sref{ss:backwardanalysis}) to identify the set of code regions reachable from each target. 
Then, for each pair of targets \( (t_i, t_j) \), \sys compares their respective reachable sets \( R_i \) and \( R_j \). If \( R_i \subset R_j \), \( t_i \) is marked as post-dominated by \( t_j \); conversely, if \( R_j \subset R_i \), \( t_j \) is post-dominated by \( t_i \) (lines 4 to 14).
Once \sys identifies all pairwise post-dominance relationships, \sys merges the targets according to their dominator and selects a single representative from each group (lines 15 to 19). This process eliminates redundant targets whose coverage is subsumed by others while preserving all reachable paths relevant to vulnerable targets.

\subsection{Target-specific Path Pruning}
\label{ss:DesignModule2}

After identifying vulnerable targets in~\Sref{ss:DesignModule1}, the directed fuzzing should traverse all possible execution paths leading to each target.
However, among the many paths that can reach each target at the graph level (\eg call graph), traversing paths that cannot trigger bugs or are not actually reachable undermines the effectiveness of directed fuzzing.
Thus, accurate pruning of these paths in \rust applications is essential.
Accordingly, \sys proposes a target-specific path pruning analysis tailored to \rust specific memory safety bug type.

\subsubsection{Target Classification and Call-site Remapping}
\label{ss:target-classificatoin-remapping}
To identify the code regions that must be visited to trigger bugs related to each target, \sys first leverages backward reachability analysis to identify all reachable basic blocks from each target.
However, existing reachability analysis techniques ~\cite{srivastava2022one, huang2022beacon} are not directly applicable in \rust applications because their path pruning shows high under-pruning ratios (\eg 51\%) and eventually causes a DGF to direct its fuzzing run toward paths that are irrelevant to the target.

\begin{algorithm}[t]
\caption{Dynamic Fair Toggling Fuzzing}
\label{algo:algorithm2}

\begin{algorithmic}[1]

\STATE \textbf{Input:}
Program \textit{Prog}, Initial seeds $S$,
Target set $T=\{t_1,\ldots,t_n\}$,
Pruning info $P=\{P[t]\mid t\in T\}$

\STATE \textbf{Output:}
Potential memory bugs

\STATE
\textbf{function} \textsc{RustGoFuzz}$(\mathit{Prog},S,T,P)$

\STATE
\hspace{\algorithmicindent}
\textbf{for all} $t\in T$ \textbf{do}

\STATE
\hspace{\algorithmicindent}
\hspace{\algorithmicindent}
$\mathit{MultiQueue}[t]
 \gets \operatorname{InitializeQueue}(S)$

\STATE
\hspace{\algorithmicindent}
\hspace{\algorithmicindent}
$\mathit{MultiBitmap}[t]
 \gets \operatorname{CreateBitmap}()$

\STATE
\hspace{\algorithmicindent}
\hspace{\algorithmicindent}
$\mathit{TSMState}[t]
 \gets \operatorname{InitializeTSM}(P[t])$

\STATE
\hspace{\algorithmicindent}
\textbf{end for}

\STATE
\hspace{\algorithmicindent}
$t_{cur}\gets\operatorname{InitTarget}(T)$

\STATE
\hspace{\algorithmicindent}
$\mathit{set}(
  \mathit{TSMState}[t_{cur}],
  \mathit{MultiBitmap}[t_{cur}],
  \mathit{MultiQueue}[t_{cur}])$

\STATE
\hspace{\algorithmicindent}
$\mathit{seed}
 \gets \operatorname{SelectInput}
       (\mathit{MultiQueue}[t_{cur}])$

\STATE
\hspace{\algorithmicindent}
\textbf{while true do}

\STATE
\hspace{\algorithmicindent}
\hspace{\algorithmicindent}
$\mathit{skip\_fuzz}
 \gets \operatorname{FuzzOne}(\mathit{Prog},\mathit{seed})$

\STATE
\hspace{\algorithmicindent}
\hspace{\algorithmicindent}
\textbf{if not} $\mathit{skip\_fuzz}$ \textbf{then}

\STATE
\hspace{\algorithmicindent}
\hspace{\algorithmicindent}
\hspace{\algorithmicindent}
$t_{cur}\gets\operatorname{NextTarget}(T)$

\STATE
\hspace{\algorithmicindent}
\hspace{\algorithmicindent}
\hspace{\algorithmicindent}
$\mathit{seed}
 \gets \operatorname{SelectInput}
       (\mathit{MultiQueue}[t_{cur}])$

\STATE
\hspace{\algorithmicindent}
\hspace{\algorithmicindent}
\hspace{\algorithmicindent}
$\mathit{update}(
  \mathit{TSMState}[t_{cur}],
  \mathit{MultiBitmap}[t_{cur}],$

\STATE
\hspace{\algorithmicindent}
$\mathit{MultiQueue}[t_{cur}])$

\STATE
\hspace{\algorithmicindent}
\hspace{\algorithmicindent}
\textbf{end if}

\STATE
\hspace{\algorithmicindent}
\textbf{end while}

\STATE
\textbf{end function}

\end{algorithmic}
\end{algorithm}

This issue mainly occurs due to the same \rust-specific features discussed in \Sref{ss:FPProblem}. In particular, the pervasive use of standard library functions such as \cc{drop}, which previously complicated taint analysis, similarly hinders precise reachability analysis by introducing numerous irrelevant paths, leading to under-pruning. 
This is because standard library functions are typically shared across many call sites, and backward reachability analysis starting from inside the library tends to over-approximate reachability by connecting the target to all such call sites.
Therefore, \sys remaps such targets to their corresponding call sites in non-standard library code.
Specifically, when a target resides inside the \rust standard library (\eg \cc{std::ptr::write}), \sys redirects the analysis from internal library to the corresponding invocation points (\eg \cc{call std::ptr::write}), as shown in \Fref{fig:path-pruning-analysis} (b).

\subsubsection{Precondition-aware Path Pruning}
\label{ss:backwardanalysis}
After \Sref{ss:target-classificatoin-remapping}, \sys obtains a finalized set of vulnerable targets, including remapped call-site targets for those originally located within the standard library.
\sys then performs path pruning to eliminate paths that are guaranteed not to satisfy the necessary conditions for triggering a memory safety bug, preventing wasted effort on exploring such paths.
To avoid false negatives (\ie pruning paths that may trigger bugs), \sys preserves all execution paths reachable from the program entry point (main) and applies pruning only in cases where memory bugs cannot occur or when eliminating infeasible paths through standard library abstraction (\Sref{ss:target-classificatoin-remapping}).
Therefore, all executable paths that can potentially trigger bugs are preserved without pruning. Accordingly, even in complex bug scenarios involving global values and shared buffers, all paths that are reachable to the actual bug-triggering targets are preserved, allowing bugs to be triggered without false negatives.

More specifically, when the target is a raw pointer dereference site, \sys does not apply any pruning, because spatial memory safety bugs can occur through raw pointer dereferences without requiring specific preconditions (\eg a dangling pointer in use-after-free). As a result, when a raw pointer dereference is the target, all related paths are preserved. 
In contrast, reference sites can only lead to temporal memory safety bugs (not spatial memory safety bugs), since Rust’s built-in bounds checks—even in unsafe code—prevent spatial memory safety bugs (as mentioned in~\Sref{ss:memorybugsinrust}). Accordingly, when the target is a safe reference site, \sys carefully applies precondition-aware pruning.

As mentioned above, \sys performs no pruning for spatial memory bug candidates. However, for temporal memory safety bugs, pruning is selectively and safely applied. In temporal memory safety bugs, although a vulnerable target may be reachable through multiple paths, only a subset of those paths satisfies the necessary bug-triggering preconditions.
\sys therefore prunes targets paths that cannot exhibit the certain sequences of memory operations, such as \cc{allocation} $\rightarrow$ \cc{free} $\rightarrow$ \cc{dereference} for use-after-free and \cc{allocation} $\rightarrow$ \cc{free} $\rightarrow$ \cc{free} for double-free.

To identify such memory operation sequences, \sys leverages the results of the inter-procedural taint analysis conducted in the \Sref{ss:DesignModule1} phase.
In that phase, if a target is identified as a possible target for a temporal memory safety bug, \sys tracks and records the memory-related operations (\eg allocation, deallocation, and dereference) encountered during taint propagation that corresponds to the required bug-triggering sequences. 
Later, when performing backward reachability analysis from the temporal memory safety bug targets, \sys utilizes this pre-collected information to examine each reachable path, verifying whether it includes the specific memory operation patterns necessary to manifest the bug.
If certain paths do not satisfy these bug conditions, \sys prunes them, since they cannot trigger temporal memory safety bugs, as shown in ~\Fref{fig:path-pruning-analysis} (c-2).

Additionally, to detect temporal bugs without false negatives, it is necessary to support async/await functions. 
More specifically, \rust's async/await constructs introduce nontrivial lifetime interactions that can affect the identification of temporal memory safety bug targets. \sys includes a concurrency-safe mode to handle this effect. In this mode, \sys conservatively and automatically detects such functions by demangling LLVM IR names and matching patterns such as \cc{GenFuture} and \cc{from\_generator}, and disables pruning for code involving raw pointers or aliasing within these functions to avoid false negatives.

\begin{figure}[t]
    \centering
    \includegraphics[width=0.44\textwidth]{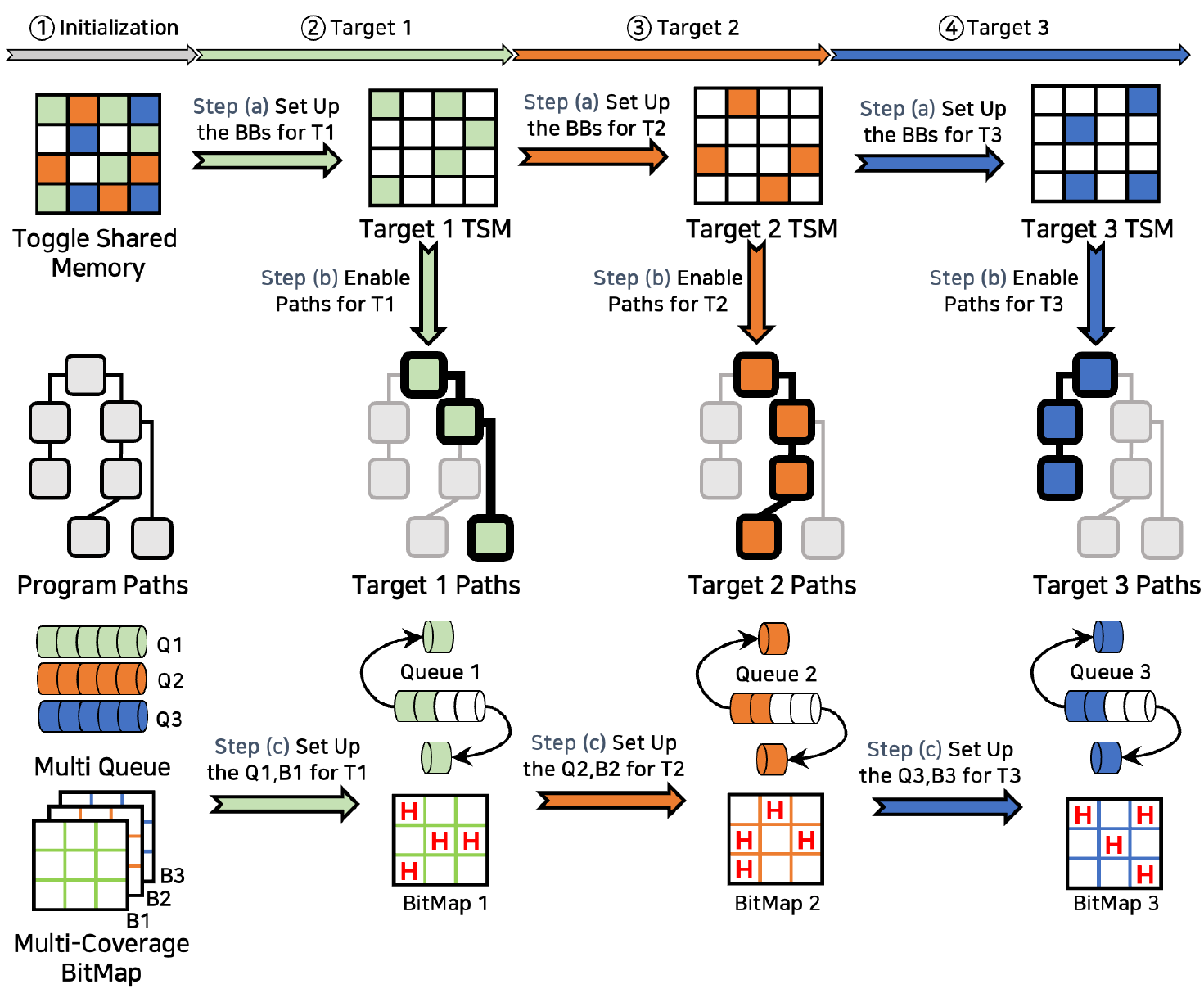}
    \caption{Overview of the fuzzing phase in dynamic toggling fuzzing with a multi-queue, multi-coverage scheme.}
    \label{fig:multi-queue}
\end{figure}

\subsection{Dynamic Fair Toggling Fuzzing}
\label{ss:DesignModule3}
As mentioned earlier, \sys identifies and optimizes vulnerable targets (\Sref{ss:DesignModule1}) and applies path pruning analysis to eliminate irrelevant code regions for each identified target (\Sref{ss:DesignModule2}). In the next step, a key challenge is how to guide fuzzing toward each selected target independently and efficiently, without interference among multiple targets.
For this, \sys introduces a fuzzing strategy that dynamically prunes the irrelevant paths (computed in \Sref{ss:DesignModule2}), thereby assigning independent directionality to each target.
\sys proposes dedicated data structures such as multi-queue and multi-coverage bitmaps to ensure fuzzing remains both target-specific and interference-free. 

\Aref{algo:algorithm2} illustrates \sys's fuzzing procedure. In this subsection, we describe two key components: dynamic toggle fuzzing (lines 12-20), which dynamically manages path pruning analysis for each target to maintain independent fuzzing direction, and the multi-queue and multi-coverage scheme (lines 4-11 and 17), which manages input queues and coverage bitmaps separately for each target to enable target-specific fuzzing without interference.

\subsubsection{Dynamic Toggle Fuzzing}
\label{ss:dynamic-pruning}
To effectively test each bug-prone target identified through \sys analysis (\eg post-dominance analysis), the fuzzer need to avoid exploring code regions unrelated to the selected target. For this, \sys proposes a dynamic toggling approach that dynamically prunes paths unrelated to the currently focused target.

The fuzzer and fuzzed target program share a portion of memory to monitor the target program's execution path. Traditionally, this shared memory is used to collect coverage information from the pre-instrumented target application during runtime. \sys extends this mechanism by introducing pruning instrumentation and additional \sys's Toggle Shared Memory (TSM) to dynamically apply target-specific pruning information. 
To enable this, \sys and the target application utilize a shared TSM that determines which basic blocks should be treated as reachable (\ie need to be visited during testing) during fuzzing execution. If execution reaches an inactive basic block, \sys preemptively terminates the execution via instrumented \cc{exit} code.

Before applying dynamic toggling, \sys first determines which target to fuzz. 
When a specific target is selected for fuzzing, inputs are selected from that target's dedicated queue. 
For this, \sys manages independent input queues for each target through the Multi-Queue scheme (\Sref{ss:multi-queue}).
Initially, all queues are populated with the same initial seeds. During fuzzing, inputs that discover new coverage along a target's paths are added to that target's queue.
After preparing the initial inputs, \sys applies a set of input mutation operators (\eg bit-flip, havoc, insertion, and deletion).
Only after this fuzzing process completes successfully (\ie \cc{fuzz\_one} returns \cc{0}), \sys moves on to the next target in a round-robin manner. This ensures that each target receives a equal fuzzing budget, preventing bias toward easily reachable targets.

As illustrated in \Fref{fig:multi-queue}, \sys dynamically updates the set of executable basic blocks each time the target is toggled.
Specifically, as shown in \Fref{fig:multi-queue}, at Step (a) of each target phase, the TSM is updated at Step (b) to enable only the basic blocks relevant to the currently scheduled target.
For instance, when fuzzing Target 1, the \sys activates only the green-highlighted paths (Target 1 TSM), while deactivating all unrelated paths. When the \sys switches to Target 2, these paths are reconfigured based on Target 2 TSM, which activates a completely different subset of the program.
This toggling of reachable paths ensures that only meaningful code regions are explored for each target, significantly reducing overhead from irrelevant paths.
Importantly, this pruning logic is entirely dynamic, meaning that the PUT remains unchanged, while the TSM is updated in real-time to reflect the currently active target.
As a result, \sys performs multiple independent fuzzing tasks in a single unified fuzzing loop, avoiding the need for separate processes or static configurations for each target.

\subsubsection{Multi-Queue and Multi-Coverage Scheme}
\label{ss:multi-queue}
Despite fair target selection in \Sref{ss:dynamic-pruning}, this approach alone cannot ensure fair testing across all targets because shared data structures hinder fair and focused fuzzing.

\noindent \textbf{Multi-Queue Scheme.} The main reason is that most existing fuzzers rely on a single input queue to distribute fuzzing energy across all targets, making it difficult to determine which input was generated for which target. Moreover, inputs for easily reachable targets appear in the queue more frequently, causing the fuzzer to focus on testing those targets. This leads to an unfair distribution of fuzzing energy, causing fuzzing starvation on hard-to-reach targets.

To resolve the above problem, \sys introduces a Multi-Queue Scheme that maintains multiple separate target-specific input queues, as shown in \Fref{fig:multi-queue}-(c). Those queues store all the seed inputs only for specific targets. When one target is scheduled to be fuzzed, \sys automatically switches to its target-specific input queue. This enables \sys to perform target-specific fuzzing without interference from other targets.

\noindent \textbf{Multi-Coverage Bitmap.} Fuzzers use a coverage map in the form of a global bitmap that records all execution paths exercised at least once by inputs in the corpus, thereby providing a global view of program coverage. However, using a single coverage map for all targets imposes a fundamental limitation: coverage data cannot be distinguished per target. It is insufficient for \sys because it must track the executed paths for every target. Otherwise, \sys cannot figure out which paths are executed for each target due to the overlapped coverage results of different targets.

To address this issue, as shown in \Fref{fig:multi-queue}-(c), \sys creates and manages an independent coverage bitmap for each target point. For every target, \sys initializes a distinct coverage map and when a specific target is scheduled to be fuzzed, \sys uses its specific bitmap to record target-specific coverages.

\section{Implementation}

\begin{table}[t]
  \centering
  \caption{Compared Fuzzer and Fuzz Driver Generator Approach List.}
  \label{tbl:comparison-fuzzer}
  \resizebox{1.0\columnwidth}{!}{%
\begin{tabular}{l|c|c}
    \toprule
    \textbf{Fuzzer}  & \textbf{Category}  & \textbf{Description}     \\
    \midrule
    \textbf{AFLGO~\cite{bohme2017directed}} 
    & \multicolumn{1}{c|}{Distance-based} 
    & \multicolumn{1}{c}{Distance-guided seed scheduling} \\ \cline{1-3}
    \textbf{WindRanger~\cite{du2022windranger}} 
    & \multicolumn{1}{c|}{Distance-based} 
    & \multicolumn{1}{c}{Deviation-aware distance estimation} \\ \cline{1-3}
    \textbf{FishFuzz~\cite{zheng2023fishfuzz}} 
    & \multicolumn{1}{c|}{Multi-target-based} 
    & \multicolumn{1}{c}{Sanitizer-based target  prioritization} \\ \cline{1-3}
    \textbf{AFLRUN~\cite{rong2024toward}} 
    & \multicolumn{1}{c|}{Multi-target-based} 
    & \multicolumn{1}{c}{Critical blocks prioritization} \\ \cline{1-3}
    \textbf{Lyso~\cite{lyso}} 
    & \multicolumn{1}{c|}{Multi-target-based} 
    & \multicolumn{1}{c}{Semantic execution-step guidance} \\ \cline{1-3}
    \textbf{PanicKiller~\cite{panickiller}} 
    & \multicolumn{1}{c|}{Multi-target-based}
    & \multicolumn{1}{c}{Unsafe-focused panic bypassing} \\ \cline{1-3}
    \textbf{AFL~\cite{web:AFL}} 
    & \multicolumn{1}{c|}{Coverage-based} 
    & \multicolumn{1}{c}{Global coverage maximization} \\ \cline{1-3}
    \textbf{AFL++~\cite{web:AFL++}} 
    & \multicolumn{1}{c|}{Coverage-based} 
    & \multicolumn{1}{c}{AFL with multiple optimizations} \\  \cline{1-3}
    \textbf{RPG~\cite{xu2024rpg}} 
    & \multicolumn{1}{c|}{Fuzz Driver Generator} 
    & \multicolumn{1}{c}{Pool-based fuzz driver generator} \\ \cline{1-3}
    \textbf{RUG~\cite{rug}} 
    & \multicolumn{1}{c|}{Fuzz Driver Generator} 
    & \multicolumn{1}{c}{LLM-based fuzz driver generator} \\ \cline{1-3}
    \textbf{deepSURF~\cite{deepsurf}} 
    & \multicolumn{1}{c|}{Fuzz Driver Generator} 
    & \multicolumn{1}{c}{LLM-based fuzz driver generator} \\ \cline{1-3}
    \bottomrule
\end{tabular}
}

\end{table}

\sys is implemented on top of \cc{rustc-1.64-dev} and AFL++-v4.0.0c.
It utilizes \cc{LLVM\,14.0.6} to instrument both coverage and sanitizers and \cc{SVF\,2.4} for static analysis.
Our implementation consists of about 13K LOC in total (mainly in the Rust compiler, SVF, and AFL++).

\noindent \textbf{Targets Identification and Optimization.}
To automatically identify vulnerable codes for memory bugs, we first identify raw pointers in target \rust applications.
For this, we modify the \cc{codegen-ssa} and \cc{codegen-llvm} components of the \rust compiler (\cc{rustc-1.64\allowbreak.0}) to perform a type-matching analysis on all statements and terminators during the \cc{HIR}/\cc{MIR} phase.
We then leverage \cc{SVF\,2.4} to automatically identify all memory-bug-prone targets related to these raw pointers.

To achieve this, we implement our taint analysis engine by integrating the Sparse Value Flow Graph (SVFG)~\cite{sui2016svf} with inter-procedural control-flow sensitivity. Specifically, we use SVFG to track fine-grained taint propagation, and augment it with ICFG edges to accurately model control-flow across function boundaries. When encountering standard library calls, our implementation skips SVFG traversal inside the library functions. Instead, it connects the tainted arguments directly to the return values using ICFG edges. 
Furthermore, to eliminate redundant analysis targets, \sys builds upon post-dominator analysis provided by \cc{LLVM\,14.0.6v} to merge target locations that exhibit post-dominance relationships across function boundaries.

\noindent \textbf{Dynamic Fair Toggling Fuzzing.}
\sys performs path pruning instrumentation based on \cc{LLVM\,14.0.6v}.
During fuzzing, it maintains target-specific reachable path information in a toggle shared memory, which is periodically updated.
This toggle shared memory (\eg \cc{TSM}) is combined with pruning instrumentation in each basic block, using code like \cc{ if (is\_pruned(TSM[cur\_bb])) { exit();}}, enabling early termination.
Additionally, \sys extends AFL++ by managing an independent coverage bitmap and input queue for each target.
Whenever the target is updated (\ie \cc{fuzz\_one} returns \cc{0}), \sys switches to the corresponding coverage bitmap and input queue, allowing target-specific fuzzing.

\section{Evaluation}
In this section, we evaluate \sys across seven aspects: effectiveness of target optimization (\Sref{ss:target-optimization}), target path pruning (\Sref{ss:target-pruning}), dynamic fair toggling (\Sref{ss:effectiveness-of-dynamic-toggling}), effectiveness of \sys (\Sref{ss:effectiveness-of-rustgo}), ablation study (\Sref{ss:ablation-study}), comparison with fuzz driver generator (\Sref{ss:comparison-with-fuzz-target-generator}), and new vulnerability detection (\Sref{ss:new-vulnerability-discovery}).

\begin{table}[t]
  \centering
  \caption{Listing of variants for fine-grained ablation study.
  \textcolor{red}{\faTimes} symbol indicates that the corresponding
  component is not applied, while
  \textcolor{green}{\faCheck} symbol denotes that it is applied.
  Each variant incrementally adds components of \sys to analyze their
  individual and cumulative impact on performance.}
  \label{tbl:ablation-module}

\begingroup

\renewcommand{\arraystretch}{1.1}
\setlength{\tabcolsep}{2pt}

\resizebox{\linewidth}{!}{%
\begin{tabular}{
  >{\centering\arraybackslash}m{4.2cm}|
  >{\centering\arraybackslash}m{1.5cm}|
  >{\centering\arraybackslash}m{1.5cm}|
  >{\centering\arraybackslash}m{1.5cm}|
  >{\centering\arraybackslash}m{1.5cm}|
  >{\centering\arraybackslash}m{1.5cm}|
  >{\centering\arraybackslash}m{1.5cm}
}
\toprule

\textbf{Configurations for Ablation Study}
& \multicolumn{1}{m{1.5cm}|}{
    \centering\rustgodissafe}
& \multicolumn{1}{m{1.5cm}|}{
    \centering\textsc{RG}\\\textsc{-base}}
& \multicolumn{1}{m{1.5cm}|}{
    \centering\textsc{RG}\\\textsc{-opt}}
& \multicolumn{1}{m{1.5cm}|}{
    \centering\textsc{RG}\\\textsc{-prun}}
& \multicolumn{1}{m{1.5cm}|}{
    \centering\textsc{RG}\\\textsc{-sync}}
& \multicolumn{1}{m{1.5cm}}{
    \centering\sys}
\\
\midrule\midrule

\multicolumn{1}{m{4.2cm}|}{
  \centering
  Coverage Feedback\\
  from Safe Code Region}
& \textcolor{red}{\faTimes}
& \textcolor{green}{\faCheck}
& \textcolor{green}{\faCheck}
& \textcolor{green}{\faCheck}
& \textcolor{green}{\faCheck}
& \textcolor{green}{\faCheck}
\\
\cline{1-7}

\multicolumn{1}{m{4.2cm}|}{
  \centering
  Target Identification\\
  and Optimization (\ref{ss:DesignModule1})}
& \textcolor{red}{\faTimes}
& \textcolor{red}{\faTimes}
& \textcolor{green}{\faCheck}
& \textcolor{green}{\faCheck}
& \textcolor{green}{\faCheck}
& \textcolor{green}{\faCheck}
\\
\cline{1-7}

\multicolumn{1}{m{4.2cm}|}{
  \centering
  Target-Specific\\
  Path Pruning (\ref{ss:DesignModule2})}
& \textcolor{red}{\faTimes}
& \textcolor{red}{\faTimes}
& \textcolor{red}{\faTimes}
& \textcolor{green}{\faCheck}
& \textcolor{green}{\faCheck}
& \textcolor{green}{\faCheck}
\\
\cline{1-7}

\multicolumn{1}{m{4.2cm}|}{
  \centering
  Async/Await Function-aware\\
  \mbox{Pruning~(\ref{ss:DesignModule2})}}
& \textcolor{red}{\faTimes}
& \textcolor{red}{\faTimes}
& \textcolor{red}{\faTimes}
& \textcolor{red}{\faTimes}
& \textcolor{green}{\faCheck}
& \textcolor{green}{\faCheck}
\\
\cline{1-7}

\multicolumn{1}{m{4.2cm}|}{
  \centering
  Dynamic Fair\\
  Toggling Fuzzing (\ref{ss:DesignModule3})}
& \textcolor{red}{\faTimes}
& \textcolor{red}{\faTimes}
& \textcolor{red}{\faTimes}
& \textcolor{red}{\faTimes}
& \textcolor{red}{\faTimes}
& \textcolor{green}{\faCheck}
\\

\bottomrule
\end{tabular}%
}

\endgroup
\end{table}

\subsection{Evaluation Setup}
\label{ss:Evaluation-setup}

\subsubsection{Evaluation Environment}
Following common experimental practices in fuzzing research~\cite{klees2018evaluating}, we execute each fuzzing campaign for 24 hours and repeat each evaluation 10 times. All evaluations are conducted on a computer equipped with a 48-core Intel(R) Xeon(R) Platinum 8268 CPU @ 2.90GHz, 565GB of DDR4 RAM, and running Ubuntu 22.04.5 LTS with Linux kernel 6.8.0.

\subsubsection{Rust-specific Benchmark Preparation} 
\label{sss:target-application}
Due to the lack of a well-established Rust-specific benchmark, we aim to construct a MAGMA~\cite{hazimeh2020magma} style benchmark tailored for Rust.
However, because Rust has a relatively short history, discovering multiple reproducible real-world bugs within a single program-as MAGMA does-is difficult.
Therefore, we construct a new benchmark for Rust by closely following the strategy of MAGMA~\cite{hazimeh2020magma}.
To this end, we first select \benchnum real-world \rust applications from OSS-Fuzz~\cite{web:oss-fuzz}, prioritizing those with the highest scores based on Principal Component Analysis (PCA)~\cite{web:pca}.
We then select reproducible memory safety bugs from RUSTSEC~\cite{web:rustsec} or reported GitHub issues and manually reconstruct 15 bugs (similar to MAGMA) into each of the selected real-world applications.
To ensure that vulnerabilities are uniformly distributed within each program, we categorize functions into three groups as follows. We first fuzz each program with AFL++ for 24 hours, record the time when each function is initially reached, and rank the functions accordingly. Based on these rankings, we divide the functions into three groups according to their initial execution times.
For each category, five internal targets are randomly selected, and reproducible memory safety bugs are constructed at those locations.
In addition to our constructed benchmark, we also include the deepSURF benchmark, consisting of 26 independently discovered real-world memory safety bugs from diverse \rust applications, to further evaluate \sys on real-world vulnerabilities.

\begin{table}[t]
    \centering
    \caption{Target Optimization Analysis Results}
    \renewcommand{\arraystretch}{0.75}
\small
\resizebox{0.95\columnwidth}{!}{%
\begin{tabular}{
    >{\centering\arraybackslash}m{3.0cm}|
    >{\centering\arraybackslash}m{3.0cm}|
    >{\centering\arraybackslash}m{3.0cm}
}
\toprule
\textbf{Programs} & \textbf{Init Targets (\#)} & \textbf{Optimized Targets (\#)} \\
\midrule\midrule
\textbf{nom} & 15 & 1 \\ 

\midrule
\textbf{json} & 15 & 3 \\ 

\midrule
\textbf{serde-yaml} & 15 & 3 \\ 

\midrule
\textbf{unicode-process} & 15 & 3 \\

\midrule
\textbf{unicode-streaming} & 15 & 1 \\ 

\midrule
\textbf{data-encoding} & 15 & 2 \\ 

\midrule
\textbf{chrono} & 15 & 4 \\ 

\midrule
\textbf{mqtt-broker} & 15 & 2 \\

\midrule
\textbf{qcms} & 15 & 3 \\ 

\midrule
\textbf{httparse} & 15 & 4 \\ 

\midrule
\textbf{quick-xml} & 15 & 3 \\ 

\midrule
\textbf{lz4\_flex} & 15 & 1 \\ 

\midrule
\textbf{gimli}& 15 & 1 \\ 
\midrule\midrule
\textbf{Average} & 15 & 2.38 \\
\bottomrule
\end{tabular}
}

    \label{tbl:target-optimization}
\end{table}

\subsubsection{Fuzzers for Comparison}
We evaluate \sys in comparison with the eleven existing fuzzers in~\Tref{tbl:comparison-fuzzer} by fuzzing the \benchnum chosen benchmark applications. Those fuzzers include two coverage-based fuzzers (AFL~\cite{web:AFL}, AFL++~\cite{web:AFL++}) and six state-of-the-art directed fuzzers (AFLGo~\cite{bohme2017directed}, WindRanger~\cite{du2022windranger}, FishFuzz~\cite{zheng2023fishfuzz}, AFLRUN~\cite{rong2024toward}, Lyso~\cite{lyso}, and PanicKiller~\cite{panickiller}) and three \rust fuzz driver generators (RPG~\cite{xu2024rpg}, RUG~\cite{rug}, and deepSURF~\cite{deepsurf}).
We choose AFL and AFL++ as the two most widely used coverage-guided fuzzers to show that our approach (directed to specific regions) is more efficient than their approach (simply maximizing code coverage).
Meanwhile, we select AFLGo and WindRanger as distance-based directed fuzzers to show that \sys's approach is more efficient in reaching target regions.

We further compare \sys with multi-target directed fuzzers, including FishFuzz, which prioritizes sanitizer-reported targets, AFLRUN, which features target path diversity metric and fair energy assignment, Lyso, a multi-target, multi-step fuzzer guided by semantic paths and alarm correlations, and PanicKiller, which prioritizes unsafe code regions with fewer runtime safety checks and mutates inputs to bypass panic assertions.
We also extend comparison to three state-of-the-art fuzz target generators (RPG, RUG, and deepSURF).

\begin{table}[t]
    \centering
    \caption{Comparison between the target pruning rates of \rustgobase, \rustgoopt, \rustgoprun, \rustgosync, and \sys. \textbf{STD} and \textbf{NS} indicate the number of basic blocks in the \rust standard library and non-standard (user-defined) code, respectively, \textbf{Reach} denotes the number of reachable basic blocks from each target, and \textbf{Rate} is the percentage of pruned basic blocks relative to the total number of basic blocks.}
    \renewcommand{\arraystretch}{1.5}
\resizebox{1.0\columnwidth}{!}{%
\begin{tabular}{>{\centering\arraybackslash}m{3.2cm}|r|r|r|
r|r|
r|r|
r|r}
    \toprule
    \multirow{2}{*}{\textbf{Benchmark}}  
    & \multicolumn{3}{c|}{\textbf{Program}} 
    & \multicolumn{2}{c|}{\makecell{\textsc{RG}\\\textsc{-base}}} 
    & \multicolumn{2}{c|}{\makecell{\textsc{RG}\\\textsc{-opt}}} 
    & \multicolumn{2}{c}{\makecell{\textsc{RG}\textsc{-prun} \textbackslash \\ \textsc{RG}\textsc{-Sync} \textbackslash \\ \sys}} \\
    \cline{2-10} 
    & \makecell{Total\\BB (\#)} & \makecell{STD\\BB (\#)} & \makecell{NS\\BB (\#)} 
    & \makecell{Reach\\(\#)} & \makecell{Rate\\(\%)} 
    & \makecell{Reach\\(\#)} & \makecell{Rate\\(\%)} 
    & \makecell{Reach\\(\#)} & \makecell{Rate\\(\%)} \\
    \midrule\midrule
    \textbf{nom} & 39,751 & 25,182 & 14,569 & 16,078 & 59.55 & 16,075 & 59.56 & 8,871 & 77.68 \\ \hline
    \textbf{json} & 38,450 & 24,770 & 13,680 & 18,447 & 52.02 & 12,865 & 66.54 & 8,005 & 79.18 \\ \hline
    \textbf{serde-yaml} & 51,282 & 34,323 & 16,959 & 28,067 & 45.27 & 25,864 & 49.57  & 10,784 & 78.97 \\ \hline
    \textbf{unicode-process} & 32,299 & 20,838 & 11,461 & 16,908 & 47.65 & 16,896 & 47.69  & 4,131 & 87.21 \\ \hline
    \textbf{unicode-streaming} & 32,113 & 20,843 & 11,270 & 16,687 & 48.04 & 16,683 & 48.05  & 7,410 & 76.93 \\ \hline
    \textbf{data-encoding} & 47,107 & 22,388 & 24,719 & 31,673 & 32.76 & 31,670 & 32.77 & 18,354 & 61.04 \\ \hline
    \textbf{chrono} & 42,124 & 25,137 & 16,987 & 21,005 & 50.14 & 20,265 & 51.89 & 10,719 & 74.55 \\ \hline
    \textbf{mqtt-broker} & 49,264 & 32,314 & 16,950 & 19,239 & 60.95 & 19,236 & 60.95 & 9,059 & 81.61 \\ \hline
    \textbf{qcms} & 40,155 & 23,031 & 17,124 & 20,516 & 48.91 & 20,250 & 49.57 & 3,790 & 90.56 \\ \hline
    \textbf{httparse} & 31,020 & 19,218 & 11,802 & 15,528 & 49.94 & 15,405 & 50.34 & 6,878 & 77.83 \\ \hline
    \textbf{quick-xml} & 34,774 & 22,469 & 12,305 & 17,482 & 49.73 & 17,191 & 50.56 & 7,011 & 79.84 \\ \hline
    \textbf{lz4\_flex} & 34,735 & 22,100 & 12,635 & 16,180 & 53.42  & 16,177 & 53.43 & 8,820 & 74.61 \\ \hline
    \textbf{gimli} & 50,204 & 31,394 & 18,810 & 17,058 & 66.02 & 17,055 & 66.03 & 9,827 & 80.43 \\ 
    \midrule\midrule
    
    \textbf{Average} 
    & \totalbb
    & \stdbb
    & \nonstdbb
    & \rustgobasebb
    & \rustgobasepr
    & \rustgooptbb
    & \rustgooptpr
    & \rustgobb
    & \rustgopr \\
    \bottomrule

\end{tabular}
}

    \label{tbl:target-pruning}
\end{table}

Among the evaluated fuzzers, afl.rs (based on AFL++)~\cite{web:afl.rs}, PanicKiller, RPG, RUG, and deepSURF are only compatible with the \rust ecosystem.
Therefore, we port the remaining six fuzzers supporting only C/C++ program fuzzing to work with our \rust programs.
Note that all these ported DGF fuzzers are based on the LLVM framework. Since Rust provides intermediate compilation output in the form of LLVM IR, existing LLVM-based DGF fuzzers can be applied without modifying their logic. Therefore, we do not modify the logic of any DGF fuzzer. We only configure the LLVM compiler with appropriate LLVM options to ensure that each DGF pass is applied to \rust code. 
In the case of FishFuzz, we also preserve its logic, but for a fair comparison, we update only its target location from the original sanitizer instrumentation to our designated targets.

Additionally, to evaluate AFLGo, we use its multi-target-directed fuzzing mode, and apply the same configuration to WindRanger, which is built on top of AFLGo.
In the case of PanicKiller, the source code is not publicly available at the time of writing, and the authors response that there are no plans to release the code.
Therefore, to enable comparison with \sys, we re-implement PanicKiller's core unsafe-region prioritization component following the design described in the paper. Specifically, as PanicKiller adopts the same unsafe code identification strategy as ERASan~\cite{min2024erasan}, we identify unsafe code regions using ERASan’s open-sourced analysis. We then re-implement PanicKiller's directionality mechanism by prioritizing inputs that reach these unsafe regions, assigning them higher selection priority to drive further mutations toward unsafe code.
Note that we cannot evaluate with other existing directed fuzzers~\cite{huang2022beacon,srivastava2022one,luo2023selectfuzz,huang2024titan} because their compiler versions are too low (\eg LLVM 4) to properly compile recent \rust applications. 
Instead, we compare \sys with Lyso, a more recent DGF that has been shown to outperform these excluded approaches.

To ensure a fair comparison across all directed fuzzers, we manually assign the same set of target locations to each fuzzer, even though \sys is capable of automatically identifying potentially vulnerable targets.
More specifically, we designate the locations where the injected vulnerabilities are triggered as the target locations for each fuzzer.
Although FishFuzz originally performs fuzzing directed to sanitizer checks, we customize it to use the same predefined target locations to ensure consistent evaluations across all the directed fuzzers.
However, for PanicKiller, to ensure a fair comparison, we do not fix target locations, as its unsafe-region–based prioritization relies on continuous feedback from multiple unsafe regions rather than reaching predefined targets.

\renewcommand{\arraystretch}{0.8} 
\begin{table*}[t]
  \centering
  \caption{The average bug exposure time of \sys and the compared fuzzers. Rate represents the arithmetic mean of the per-bug detection time ratios of a fuzzer to those of RustGo. Total Time Ratio (TTR) represents the ratio of the total detection time of a fuzzer to that of RustGo. } 
  \label{tbl:performance}
\renewcommand{\arraystretch}{1.2}
\huge
\resizebox{\textwidth}{!}{%
\begin{tabular}{
>{\centering\arraybackslash}m{4.5cm}| 
>{\raggedleft\arraybackslash}m{2.0cm}|
>{\centering\arraybackslash}m{1.6cm}|
>{\centering\arraybackslash}m{1.6cm}|
>{\raggedleft\arraybackslash}m{2.0cm}|
>{\centering\arraybackslash}m{1.6cm}|
>{\centering\arraybackslash}m{1.6cm}|
>{\raggedleft\arraybackslash}m{2.0cm}|
>{\centering\arraybackslash}m{1.6cm}|
>{\centering\arraybackslash}m{1.6cm}|
>{\raggedleft\arraybackslash}m{2.0cm}|
>{\centering\arraybackslash}m{1.6cm}|
>{\centering\arraybackslash}m{1.6cm}|
>{\raggedleft\arraybackslash}m{2.0cm}|
>{\centering\arraybackslash}m{1.6cm}|
>{\centering\arraybackslash}m{1.6cm}|
>{\raggedleft\arraybackslash}m{2.0cm}|
>{\centering\arraybackslash}m{1.6cm}|
>{\centering\arraybackslash}m{1.6cm}|
>{\raggedleft\arraybackslash}m{2.0cm}|
>{\centering\arraybackslash}m{1.6cm}|
>{\centering\arraybackslash}m{1.6cm}|
>{\raggedleft\arraybackslash}m{2.0cm}|
>{\centering\arraybackslash}m{1.6cm}|
>{\centering\arraybackslash}m{1.6cm}|
>{\centering\arraybackslash}m{2.0cm}
}
\toprule
\multirow{2}{*}{\textbf{Program}} 
& \multicolumn{3}{c|}{\textbf{AFL}} 
& \multicolumn{3}{c|}{\textbf{AFL++}} 
& \multicolumn{3}{c|}{\textbf{AFLGO}} 
& \multicolumn{3}{c|}{\textbf{WindRanger}} 
& \multicolumn{3}{c|}{\textbf{FishFuzz}} 
& \multicolumn{3}{c|}{\textbf{AFLRUN}} 
& \multicolumn{3}{c|}{\textbf{Lyso}} 
& \multicolumn{3}{c|}{\textbf{PanicKiller}} 
& \multicolumn{1}{c}{\textbf{RustGo}} \\
\cline{2-26}
& \centering\textbf{Time (ms)} & \textbf{Rate ($\times$)} & \textbf{TTR ($\times$)} 
& \centering\textbf{Time (ms)} & \textbf{Rate ($\times$)} & \textbf{TTR ($\times$)} 
& \centering\textbf{Time (ms)} & \textbf{Rate ($\times$)} & \textbf{TTR ($\times$)} 
& \centering\textbf{Time (ms)} & \textbf{Rate ($\times$)} & \textbf{TTR ($\times$)} 
& \centering\textbf{Time (ms)} & \textbf{Rate ($\times$)} & \textbf{TTR ($\times$)} 
& \centering\textbf{Time (ms)} & \textbf{Rate ($\times$)} & \textbf{TTR ($\times$)}
& \centering\textbf{Time (ms)} & \textbf{Rate ($\times$)} & \textbf{TTR ($\times$)}
& \centering\textbf{Time (ms)} & \textbf{Rate ($\times$)} & \textbf{TTR ($\times$)}
& \textbf{Time (ms)}  \\

\midrule\midrule
\textbf{nom} 
& 19,082 & $\times$3.69 & $\times$5.18   
& 11,362 & $\times$2.57 & $\times$3.08     
& 16,192 & $\times$2.67 & $\times$4.40 
& 35,157 & $\times$3.32 & $\times$9.54 
& 14,601 & $\times$4.23 & $\times$3.96   
& 8,198 & $\times$2.14 & $\times$2.23    
& 7,629 & $\times$2.34 & $\times$2.07
& 8,532 & $\times$2.58 & $\times$2.32
& 3,684   \\

\midrule
\textbf{json} 
& 235,315 & $\times$6.94 & $\times$4.71 
& 110,201 & $\times$2.93 & $\times$2.21 
& 655,446 & $\times$6.88 & $\times$13.13 
& 407,851 & $\times$6.17 & $\times$8.17 
& 78,445 & $\times$2.86 & $\times$1.57 
& 247,152 & $\times$2.62 & $\times$4.95
& 130,238 & $\times$3.21 & $\times$2.61
& 57,261 & $\times$1.97 & $\times$1.15
& 49,918   \\

\midrule
\textbf{serde-yaml} 
& 370,504 & $\times$2.18 & $\times$1.36 
& 253,308 & $\times$1.63 & $\times$0.93 
& 742,500 & $\times$3.50 & $\times$2.72 
& 746,812 & $\times$3.94 & $\times$2.74 
& 703,722 & $\times$9.93 & $\times$2.58 
& 391,043 & $\times$1.97 & $\times$1.43
& 662,031 & $\times$2.56 & $\times$2.42
& 358,400 & $\times$1.50 & $\times$1.31
& 273,012  \\

\midrule
\textbf{unicode-process} 
& 27,357 & $\times$2.21 & $\times$1.61  
& 37,798 & $\times$1.62 & $\times$2.22  
& 28,311 & $\times$2.32 & $\times$1.66  
& 48,507 & $\times$3.41 & $\times$2.85 
& 17,985 & $\times$2.82 & $\times$1.06  
& 24,850 & $\times$2.09 & $\times$1.46  
& 22,857 & $\times$2.05 & $\times$1.34 
& 34,232 & $\times$1.97 & $\times$2.01  
& 17,041  \\

\midrule
\textbf{unicode-streaming} 
& 168,235 & $\times$1.78 & $\times$3.03   
& 220,678 & $\times$4.67 & $\times$3.97   
& 182,109 & $\times$2.62 & $\times$3.28  
& 695,805 & $\times$7.53 & $\times$12.52 
& 193,798 & $\times$2.83 & $\times$3.49   
& 211,127 & $\times$4.67 & $\times$3.80   
& 178,225 & $\times$4.25 & $\times$3.21 
& 80,256 & $\times$1.45 & $\times$1.44
& 55,566  \\

\midrule
\textbf{data-encoding} 
& 2,667 & $\times$1.68 & $\times$2.00 
& 3,261 & $\times$1.72 & $\times$2.44 
& 7,073 & $\times$4.28 & $\times$5.30 
& 3,229 & $\times$2.25 & $\times$2.42 
& 4,404 & $\times$3.38 & $\times$3.30 
& 1,442 & $\times$1.28 & $\times$1.08 
& 3,141 & $\times$2.06 & $\times$2.35 
& 2,836 & $\times$1.89 & $\times$2.12
& 1,335  \\

\midrule
\textbf{chrono} 
& 2,159,152 & $\times$1.28 & $\times$1.27
& 1,893,340 & $\times$1.11 & $\times$1.11 
& 6,824,765 & $\times$4.03 & $\times$4.00  
& 5,959,157 & $\times$3.59 &$\times$3.50  
& 5,957,256 & $\times$3.43 & $\times$3.49  
& 1,825,604 & $\times$1.08 & $\times$1.07  
& 3,568,652 & $\times$3.46 & $\times$2.09
& 1,842,845 & $\times$1.08 & $\times$1.08
& 1,704,923  \\

\midrule
\textbf{mqtt-broker} 
& 20,105 & $\times$2.01 & $\times$1.19 
& 130,128 & $\times$4.85 & $\times$7.69  
& 31,113 & $\times$2.50 & $\times$1.84  
& 62,559 & $\times$3.03 & $\times$3.70   
& 33,176 & $\times$3.26 & $\times$1.96   
& 24,636 & $\times$1.34 & $\times$1.46 
& 28,313 & $\times$2.25 & $\times$1.67  
& 19,156 & $\times$1.39 & $\times$1.13 
& 16,912  \\

\midrule
\textbf{qcms} 
& 1,119,483 & $\times$2.69 & $\times$2.51 
& 545,303 & $\times$1.21 & $\times$1.22 
& 2,648,339 & $\times$5.14 & $\times$5.93 
& 1,739,851 & $\times$3.77 & $\times$3.90 
& 461,117 & $\times$1.52 & $\times$1.03 
& 469,364 & $\times$1.50 & $\times$1.05  
& 462,767 & $\times$1.27 & $\times$1.04 
& 854,554 & $\times$1.62 & $\times$1.91 
& 446,562  \\

\midrule
\textbf{httparse}
& 5,646,222 & $\times$9.34 & $\times$10.52     
& 3,271,086 & $\times$5.45 & $\times$6.09     
& 5,060,852 & $\times$8.48 & $\times$9.43     
& 5,671,562 & $\times$9.47 & $\times$10.56  
& 1,790,760 & $\times$3.21 & $\times$3.34   
& 1,805,783 & $\times$3.39 & $\times$3.36 
& 1,485,398 & $\times$2.52 & $\times$2.77 
& 2,056,293 & $\times$3.50 & $\times$3.83 
& 536,884  \\

\midrule
\textbf{quick-xml}
& 2,984,028 & $\times$5.78 & $\times$12.59   
& 1,755,357 & $\times$4.01 & $\times$7.40    
& 3,299,198 & $\times$3.28 & $\times$13.92   
& 4,018,965 & $\times$4.87 & $\times$16.95   
& 2,495,417 & $\times$5.08 & $\times$10.52    
& 688,241 & $\times$1.68 & $\times$2.90
& 664,142 & $\times$1.57 & $\times$2.80
& 692,224 & $\times$2.12 & $\times$2.92
& 237,095  \\

\midrule
\textbf{lz4\_flex}
& 186,262 & $\times$3.21 & $\times$1.55 
& 269,638 & $\times$3.41 & $\times$2.24 
& 202,340 & $\times$3.39 & $\times$1.68 
& 338,848 & $\times$4.53 & $\times$2.82 
& 507,893 & $\times$5.89 & $\times$4.22 
& 248,131 & $\times$2.43 & $\times$2.06 
& 308,547 & $\times$4.02 & $\times$2.57
& 355,384 & $\times$2.40 & $\times$2.96
& 120,243 \\

\midrule
\textbf{gimli}
& 1,343,662 & $\times$2.08 & $\times$2.24   
& 895,306 & $\times$1.88 & $\times$1.49   
& 2,532,443 & $\times$6.82 & $\times$4.22 
& 4,066,599 & $\times$10.16 & $\times$6.78 
& 2,507,864 & $\times$5.47 & $\times$4.18 
& 621,963 & $\times$1.18 & $\times$1.04 
& 1,737,103 & $\times$2.40 & $\times$2.90
& 1,412,436 & $\times$3.75 & $\times$2.35
& 599,871  \\

\midrule\midrule
\textbf{Average} 
& \afltime & \afl & \aflttr
& \aflpptime & \aflpp & \aflppttr 
& \aflgotime & \aflgo & \aflgottr 
& \windrangertime & \windranger & \windrangerttr 
& \fishfuzztime & \fishfuzz & \fishfuzzttr 
& \aflruntime & \aflrun & \aflrunttr   
& \lysotime & \lyso & \lysottr 
& \panickillertime & \panickiller & \panickillerttr  
& \rustgotime  \\
\bottomrule
\end{tabular}%
}

\end{table*}

\renewcommand{\arraystretch}{0.8} 
\begin{table*}[t]
  \centering
  \caption{The average bug exposure time of \sys and ablation modules. 
  }
  \label{tbl:ablation}
  \renewcommand{\arraystretch}{1.0}
\Large
\resizebox{\textwidth}{!}{%
\begin{tabular}{
>{\centering\arraybackslash}m{4.5cm}| 
>{\raggedleft\arraybackslash}m{2.0cm}|
>{\centering\arraybackslash}m{1.6cm}|
>{\centering\arraybackslash}m{1.6cm}|
>{\raggedleft\arraybackslash}m{2.0cm}|
>{\centering\arraybackslash}m{1.6cm}|
>{\centering\arraybackslash}m{1.6cm}|
>{\raggedleft\arraybackslash}m{2.0cm}|
>{\centering\arraybackslash}m{1.6cm}|
>{\centering\arraybackslash}m{1.6cm}|
>{\raggedleft\arraybackslash}m{2.0cm}|
>{\centering\arraybackslash}m{1.6cm}|
>{\centering\arraybackslash}m{1.6cm}|
>{\raggedleft\arraybackslash}m{2.0cm}|
>{\centering\arraybackslash}m{1.6cm}|
>{\centering\arraybackslash}m{1.6cm}|
>{\centering\arraybackslash}m{2.0cm}
}
\toprule
\multirow{2}{*}{\textbf{Program}} 
& \multicolumn{3}{c|}{\textbf{\rustgodissafe}} 
& \multicolumn{3}{c|}{\textbf{\rustgobase}} 
& \multicolumn{3}{c|}{\textbf{\rustgoopt}} 
& \multicolumn{3}{c|}{\textbf{\rustgoprun}} 
& \multicolumn{3}{c|}{\textbf{\rustgosync}} 
& \multicolumn{1}{c}{\textbf{\sys}} \\
\cline{2-17}
& \centering\textbf{Time (ms)} & \textbf{Rate ($\times$)} & \textbf{TTR ($\times$)}  
& \centering\textbf{Time (ms)} & \textbf{Rate ($\times$)} & \textbf{TTR ($\times$)}  
& \centering\textbf{Time (ms)} & \textbf{Rate ($\times$)} & \textbf{TTR ($\times$)}  
& \centering\textbf{Time (ms)} & \textbf{Rate ($\times$)} & \textbf{TTR ($\times$)}  
& \centering\textbf{Time (ms)} & \textbf{Rate ($\times$)} & \textbf{TTR ($\times$)}  
& \textbf{Time (ms)} \\
\midrule\midrule
\textbf{nom} 
& 50,268 & $\times$3.56 & $\times$13.65 
& 32,729   & $\times$3.89 & $\times$8.88  
& 25,011   & $\times$2.72 & $\times$6.79   
& 3,438   & $\times$1.00 & $\times$0.93  
& 3,171    & $\times$1.03 & $\times$0.86   
& 3,684    \\
\midrule
\textbf{json} 
& 72,540 & $\times$3.07 & $\times$1.45  
& 47,244 & $\times$2.34 & $\times$0.95  
& 27,919 & $\times$1.58 & $\times$0.56   
& 62,810  & $\times$1.54 & $\times$1.26   
& 53,219 & $\times$1.55 & $\times$1.07   
& 49,918   \\
\midrule
\textbf{serde-yaml} 
& 635,600 & $\times$2.75 & $\times$2.33  
& 615,300   & $\times$2.62 & $\times$2.25  
& 424,467   & $\times$1.55 & $\times$1.55 
& 435,086   & $\times$1.58 & $\times$1.59  
& 467,796   & $\times$1.59 & $\times$1.71  
& 273,012  \\
\midrule
\textbf{unicode-process} 
& 27,410 & $\times$1.54 & $\times$1.61  
& 28,685 & $\times$1.29 & $\times$1.68  
& 36,517 & $\times$1.16 & $\times$2.14 
& 19,023 & $\times$1.37 & $\times$1.12   
& 7,295  & $\times$1.36 & $\times$0.43  
& 17,041   \\
\midrule
\textbf{unicode-streaming} 
& 128,907  & $\times$1.58 & $\times$2.32  
& 178,112  & $\times$2.64 & $\times$3.21  
& 205,466  & $\times$3.95 & $\times$3.70  
& 75,175  & $\times$1.01 & $\times$1.35  
& 43,602  & $\times$1.02 & $\times$0.78  
& 55,566   \\
\midrule
\textbf{data-encoding} 
& 2,065 & $\times$2.28 & $\times$1.55 
& 4,354 & $\times$2.12 & $\times$3.26  
& 4,981 & $\times$2.07 & $\times$3.73  
& 3,274 & $\times$1.62 & $\times$2.45  
& 4,210 & $\times$1.66 & $\times$3.15  
& 1,335   \\
\midrule
\textbf{chrono} 
& 7,671,550 & $\times$4.82 & $\times$4.50  
& 5,190,191 & $\times$3.15 & $\times$3.04 
& 4,824,752 & $\times$2.95 & $\times$2.83  
& 2,730,807 & $\times$1.59 & $\times$1.60  
& 2,770,236 & $\times$1.60 & $\times$1.62  
& 1,704,923  \\
\midrule
\textbf{mqtt-broker} 
& 109,840 & $\times$3.55 & $\times$6.49  
& 128,478  & $\times$6.05 & $\times$7.60  
& 114,913  & $\times$7.14 & $\times$6.79 
& 20,416  & $\times$1.21 & $\times$1.21 
& 22,967  & $\times$1.22 & $\times$1.36  
& 16,912  \\
\midrule
\textbf{qcms} 
& 853,214 & $\times$1.70 & $\times$1.91  
& 681,287  & $\times$1.37  & $\times$1.53   
& 509,168  & $\times$1.03  & $\times$1.14  
& 454,501  & $\times$1.31  & $\times$1.02   
& 451,277 & $\times$1.29  & $\times$1.01    
& 446,562  \\
\midrule
\textbf{httparse}
& 5,181,259 & $\times$8.99 & $\times$9.65    
& 4,005,114 & $\times$6.81 & $\times$7.46    
& 2,279,298 & $\times$4.12 & $\times$4.25    
&   955,966 & $\times$1.72 & $\times$1.78    
&   896,347 & $\times$1.70 & $\times$1.67 
&   536,884  \\
\midrule
\textbf{quick-xml}
& 784,455  & $\times$2.04 & $\times$3.31  
& 1,037,071 & $\times$3.79 & $\times$4.37 
& 522,173  & $\times$2.88 & $\times$2.20   
& 265,572  & $\times$1.60 & $\times$1.12  
& 297,796   & $\times$1.61 & $\times$1.26  
& 237,095   \\
\midrule
\textbf{lz4\_flex}
& 168,912 & $\times$2.21 & $\times$1.40  
& 96,113  & $\times$1.34 & $\times$0.80  
& 110,961 & $\times$1.55 & $\times$0.92  
& 68,400  & $\times$1.03 & $\times$0.57  
& 34,915 & $\times$1.01 & $\times$0.29   
& 120,243   \\
\midrule
\textbf{gimli}
& 1,905,671 & $\times$4.39 & $\times$3.18  
& 1,032,278 & $\times$2.43 & $\times$1.72  
&   498,740 & $\times$1.08 & $\times$0.83  
&   424,595 & $\times$1.03 & $\times$0.71 
& 562,174 & $\times$1.04 & $\times$0.94  
&   599,871  \\
\midrule\midrule
\textbf{Average} 
& \rustgodissafetime & \rustgodissaferesult & \rustgodissafettr 
& \rustgobasetime    & \rustgobaseresult    & \rustgobasettr
& \rustgoopttime     & \rustgooptresult     & \rustgooptttr
& \rustgopruntime    & \rustgoprunresult    & \rustgoprunttr 
& \rustgosynctime    & \rustgosyncresult    & \rustgosyncttr
& \rustgotime         \\
\bottomrule
\end{tabular}%
}

\end{table*}

\subsubsection{Ablation Study Modes}
\label{sss:fine-grained-ablation}
We introduce six variant versions, namely \rustgodissafe, \rustgobase, \rustgoopt, \rustgoprun, \rustgosync, and \sys as shown in~\Tref{tbl:ablation-module}, to show the contributions of our respective proposed techniques.
\rustgodissafe disables coverage feedback for safe and performs fuzzing by considering only the coverage feedback from unsafe code regions.
\rustgobase disables all of the proposed techniques and relies solely on a reachability-based analysis based on the tripwiring~\cite{srivastava2022one}. 
\rustgoopt enables vulnerable target optimization (\Sref{ss:DesignModule1}) to consolidate redundant targets using post-dominance analysis.
\rustgoopt also performs the general reachability-based analysis on the optimized target locations.
\rustgoprun performs the target-specific reachability analysis (\Sref{ss:DesignModule2}).
To provide a concurrency-safe mode that supports async/await function handling, \rustgosync pre-identifies functions involving asynchronous or concurrent execution and disables pruning for code involving raw pointers or aliasing relationships within them.
\sys fully applies all the proposed techniques by including our dynamic fair toggling fuzzing with multi queues and multi-coverage bitmap (\Sref{ss:DesignModule3}).

\subsection{Effectiveness of Target Optimization}
\label{ss:target-optimization}

We evaluate how effectively \sys reduces redundant fuzzing targets by applying the post-dominance analysis to the benchmark programs, as described in~\Sref{sss:target-application}.
This evaluation measures how many redundant target points are consolidated into representative locations via post-dominance analysis across \benchnum benchmark applications, each of which initially contains 15 internal targets without target optimization.

As shown in~\Tref{tbl:target-optimization}, \sys effectively reduces the total number of targets by approximately \optimizedtargetrate through post-dominance analysis, retaining fewer than 3 out of 15 targets on average.
Notably, \sys merges all the 15 inserted target locations into a single representative target in \cc{nom}, \cc{unicode-streaming}, \cc{lz4\_flex}, and \cc{gimli} as those targets reside on the same execution path.
On the other hand, \cc{chrono} and \cc{httparse} retain the highest number of target locations (\ie 4 out of 15), as the internal targets reside on distinct execution paths branched by a \cc{match} statement at the early program execution stage. Therefore, the subsequent function calls in each branch do not have post-dominance relationships.
However, \sys still successfully consolidates about 73.33\% of all targets in \cc{chrono} and \cc{httparse}. Overall, across all benchmarks, \sys reduces \optimizedtargetrate targets on average, demonstrating the effectiveness of our optimization.

\renewcommand{\arraystretch}{1.0} 
\begin{table*}[t]
  \centering
  \caption{Results on the deepSURF benchmark for both performance evaluation and ablation study. Time denotes the bug exposure time, and Rate represents the ratio of each fuzzer's bug exposure time to that of \sys.}
  \label{tbl:deepsurf-benchmark-total}
\renewcommand{\arraystretch}{1.2}
\Huge
\resizebox{\textwidth}{!}{%
\begin{tabular}{
>{\centering\arraybackslash}m{1cm}|
>{\centering\arraybackslash}m{4cm}|
>{\raggedleft\arraybackslash}m{2.3cm}|
>{\centering\arraybackslash}m{1.7cm}|
>{\raggedleft\arraybackslash}m{2.3cm}|
>{\centering\arraybackslash}m{1.7cm}|
>{\raggedleft\arraybackslash}m{2.3cm}|
>{\centering\arraybackslash}m{1.7cm}|
>{\raggedleft\arraybackslash}m{2.3cm}|
>{\centering\arraybackslash}m{1.7cm}|
>{\raggedleft\arraybackslash}m{2.3cm}|
>{\centering\arraybackslash}m{1.7cm}|
>{\raggedleft\arraybackslash}m{2.3cm}|
>{\centering\arraybackslash}m{1.7cm}|
>{\raggedleft\arraybackslash}m{2.3cm}|
>{\centering\arraybackslash}m{1.7cm}|
>{\raggedleft\arraybackslash}m{2.3cm}|
>{\centering\arraybackslash}m{1.7cm}||
>{\raggedleft\arraybackslash}m{2.3cm}|
>{\centering\arraybackslash}m{1.7cm}|
>{\raggedleft\arraybackslash}m{2.3cm}|
>{\centering\arraybackslash}m{1.7cm}|
>{\raggedleft\arraybackslash}m{2.3cm}|
>{\centering\arraybackslash}m{1.7cm}|
>{\raggedleft\arraybackslash}m{2.3cm}|
>{\centering\arraybackslash}m{1.7cm}|
>{\raggedleft\arraybackslash}m{2.3cm}|
>{\centering\arraybackslash}m{1.7cm}||
>{\centering\arraybackslash}m{2.3cm}
}
\toprule
\multicolumn{2}{c|}{\multirow{2}{*}{\textbf{Benchmark}}}
& \multicolumn{16}{c||}{\textbf{Compared Fuzzer Approaches (\Sref{ss:effectiveness-of-rustgo})}}
& \multicolumn{10}{c||}{\textbf{Variants for ablation study (\Sref{ss:ablation-study})}}
& \multirow{2}{*}{\textbf{RustGo}} \\

\cline{3-28}

\multicolumn{2}{c|}{}
& \multicolumn{2}{c|}{\textbf{AFL}}
& \multicolumn{2}{c|}{\textbf{AFL++}}
& \multicolumn{2}{c|}{\textbf{AFLGo}}
& \multicolumn{2}{c|}{\textbf{WindRanger}}
& \multicolumn{2}{c|}{\textbf{FishFuzz}}
& \multicolumn{2}{c|}{\textbf{AFLRUN}}
& \multicolumn{2}{c|}{\textbf{Lyso}}
& \multicolumn{2}{c||}{\textbf{PanicKiller}}
& \multicolumn{2}{c|}{\textbf{\rustgodissafe}}
& \multicolumn{2}{c|}{\textbf{\rustgobase}}
& \multicolumn{2}{c|}{\textbf{\rustgoopt}}
& \multicolumn{2}{c|}{\textbf{\rustgoprun}}
& \multicolumn{2}{c||}{\textbf{\rustgosync}}
& \\ 
\cline{1-29}
\centering\textbf{\#} & \textbf{Program} 
& \centering\textbf{Time (ms)} & \textbf{Rate ($\times$)} 
& \centering\textbf{Time (ms)} & \textbf{Rate ($\times$)} 
& \centering\textbf{Time (ms)} & \textbf{Rate ($\times$)} 
& \centering\textbf{Time (ms)} & \textbf{Rate ($\times$)} 
& \centering\textbf{Time (ms)} & \textbf{Rate ($\times$)} 
& \centering\textbf{Time (ms)} & \textbf{Rate ($\times$)}
& \centering\textbf{Time (ms)} & \textbf{Rate ($\times$)}
& \centering\textbf{Time (ms)} & \textbf{Rate ($\times$)}
& \centering\textbf{Time (ms)} & \textbf{Rate ($\times$)} 
& \centering\textbf{Time (ms)} & \textbf{Rate ($\times$)} 
& \centering\textbf{Time (ms)} & \textbf{Rate ($\times$)}
& \centering\textbf{Time (ms)} & \textbf{Rate ($\times$)}
& \centering\textbf{Time (ms)} & \textbf{Rate ($\times$)}
& \textbf{Time (ms)}  \\

\midrule\midrule
1        & \textbf{algorithmica}
& 192,499  & $\times$9.41	
& 110,091  & $\times$5.38	
& 155,639  & $\times$7.60	
& 146,416  & $\times$7.15	
& 163,909  & $\times$8.01	
&  77,008  & $\times$3.76	
&  74,083  & $\times$3.62	
&  53,727  & $\times$2.63
& 74,380	& $\times$3.63	
& 55,706	& $\times$2.72	
& 31,650	& $\times$1.55	
& 30,381	& $\times$1.48	
& 30,663	& $\times$1.50	
&  20,466 \\

\midrule
2        & \textbf{toodee}
& 1,155	& $\times$7.31	
& 427	& $\times$2.70	
& 748	& $\times$4.73	
& 246	& $\times$1.56
& 1,331	& $\times$8.42	
& 678	& $\times$4.29	
& 1,545	& $\times$9.78	
& 385	& $\times$2.44	
& 1,115	& $\times$7.06	
& 797	& $\times$5.04	
& 845	& $\times$5.35	
& 589	& $\times$3.73	
& 514	& $\times$3.25	
& 158 \\

\midrule
3        & \textbf{toodee}
& 1,766	& $\times$4.09	
& 2,531	& $\times$5.86	
& 1,651	& $\times$3.82	
& 751	& $\times$1.74	
& 1,449	& $\times$3.35	
& 925	& $\times$2.14	
& 943	& $\times$2.18	
& 1,432	& $\times$3.31	
& 2,207	& $\times$5.11	
& 1,855	& $\times$4.29	
& 1,741	& $\times$4.03	
& 1,557	& $\times$3.60	
& 1,572	& $\times$3.64	
& 432 \\

\midrule
4        & \textbf{toodee}
& 1,435,835	& $\times$2.35	
& 697,613	& $\times$1.14	
& 1,275,003	& $\times$2.08	
& 783,731	& $\times$1.28	
& 1,019,367	& $\times$1.67	
& 1,634,603	& $\times$2.67	
& 635,317	& $\times$1.04	
& 653,694	& $\times$1.07	
& 687,159	& $\times$1.12	
& 846,260	& $\times$1.38	
& 718,214	& $\times$1.17	
& 591,281	& $\times$0.97	
& 591,996	& $\times$0.97	
& 611,883 \\

\midrule
5        & \textbf{toodee}
& 134,981	& $\times$2.93	
& 201,881	& $\times$4.38	
& 271,719	& $\times$5.90	
& 125,388	& $\times$2.72	
& 236,207	& $\times$5.12	
& 124,677	& $\times$2.71	
& 89,212	& $\times$1.94	
& 61,461	& $\times$1.33	
& 121,439	& $\times$2.63	
& 111,013	& $\times$2.41	
& 109,953	& $\times$2.39	
& 78,103	& $\times$1.69	
& 77,301	& $\times$1.68	
& 46,090 \\

\midrule
6        & \textbf{stack\_dst}
& 3,217,475	& $\times$3.58	
& 2,478,530	& $\times$2.76	
& 3,030,988	& $\times$3.37	
& 2,906,034	& $\times$3.23	
& 4,721,574	& $\times$5.25	
& 1,671,229	& $\times$1.86	
& 1,295,623	& $\times$1.44	
& 2,401,252	& $\times$2.67	
& 1,222,773	& $\times$1.36	
& 1,009,702	& $\times$1.12	
& 964,159	& $\times$1.07	
& 895,720	& $\times$1.00	
& 904,398	& $\times$1.01	
& 899,474 \\

\midrule
7        & \textbf{stack\_dst}
& 704,752	& $\times$1.21	
& 691,474	& $\times$1.19	
& 700,031	& $\times$1.20	
& 664,470	& $\times$1.14	
& 659,741	& $\times$1.13	
& 631,679	& $\times$1.08	
& 614,470	& $\times$1.05	
& 672,470	& $\times$1.15	
& 691,050	& $\times$1.19	
& 664,148	& $\times$1.14	
& 653,167	& $\times$1.12	
& 613,531	& $\times$1.05	
& 612,677	& $\times$1.05	
& 582,666 \\

\midrule
8        & \textbf{slice\_deque}
& 395,826	& $\times$4.61	
& 241,654	& $\times$2.81	
& 244,976	& $\times$2.85	
& 99,976	& $\times$1.16	
& 200,076	& $\times$2.33	
& 134,755	& $\times$1.57	
& 232,421	& $\times$2.71	
& 183,184	& $\times$2.13	
& 99,244	& $\times$1.16	
& 102,897	& $\times$1.20	
& 102,376	& $\times$1.19	
& 104,789	& $\times$1.22	
& 108,540	& $\times$1.26	
& 85,877 \\

\midrule
9        & \textbf{slice\_deque}
& 256,142	& $\times$11.94	
& 98,368	    & $\times$4.59	
& 272,417	& $\times$12.70	
& 403,621	& $\times$18.81	
& 371,766	& $\times$17.33	
& 108,201	& $\times$5.04	
& 123,672	& $\times$5.76	
& 124,050	& $\times$5.78	
& 168,805	& $\times$7.87	
& 146,725	& $\times$6.84	
& 96,416	& $\times$4.49	
& 44,744	& $\times$2.09	
& 48,700	& $\times$2.27	
& 21,453 \\

\midrule
10        & \textbf{slice\_deque}
& 102,599	& $\times$2.20	
& 97,764	& $\times$2.10	
& 190,319	& $\times$4.09	
& 188,612	& $\times$4.05	
& 196,878	& $\times$4.23	
& 61,633	& $\times$1.32	
& 65,478	& $\times$1.41	
& 122,015	& $\times$2.62	
& 55,184	& $\times$1.18	
& 61,814	& $\times$1.33	
& 66,414	& $\times$1.43	
& 46,725	& $\times$1.00	
& 47,019	& $\times$1.01	
& 46,582 \\

\midrule
11        & \textbf{slice\_deque}
& 9,072	    & $\times$5.20	
& 7,872	    & $\times$4.51	
& 9,395	    & $\times$5.39	
& 15,866	& $\times$9.10	
& 13,459	& $\times$7.72	
& 5,019	    & $\times$2.88	
& 1,778	    & $\times$1.02	
& 10,515	& $\times$6.03	
& 4,404	& $\times$2.53	
& 10,012	& $\times$5.74	
& 10,316	& $\times$5.92	
& 8,032	& $\times$4.61	
& 8,129	& $\times$4.66	
& 1,744 \\

\midrule
12        & \textbf{slice\_deque}
& 2,529,521	& $\times$7.81	
& 652,674	& $\times$2.02	
& 1,640,161	& $\times$5.06	
& 1,096,786	& $\times$3.39	
& 2,471,930	& $\times$7.63	
& 924,013	& $\times$2.85	
& 687,141	& $\times$2.12	
& 628,852	& $\times$1.94	
& 475,706	& $\times$1.47	
& 394,380	& $\times$1.22	
& 334,952	& $\times$1.03	
& 303,568	& $\times$0.94	
& 304,209	& $\times$0.94	
& 323,891 \\

\midrule
13        & \textbf{stackvector}
& 752,452	& $\times$3.97	
& 658,608	& $\times$3.47	
& 1,241,951	& $\times$6.55	
& 1,375,825	& $\times$7.25	
& 605,469	& $\times$3.19	
& 422,895	& $\times$2.23	
& 428,488	& $\times$2.26	
& 608,880	& $\times$3.21	
& 459,736	& $\times$2.42	
& 365,069	& $\times$1.92	
& 386,781	& $\times$2.04	
& 380,607	& $\times$2.01	
& 383,719	& $\times$2.02	
& 189,671 \\

\midrule
14        & \textbf{insert\_many}
& 1,728,754	& $\times$2.84	
& 697,166	& $\times$1.15	
& 1,586,280	& $\times$2.61	
& 748,848	& $\times$1.23	
& 837,844	& $\times$1.38	
& 695,921	& $\times$1.14	
& 935,061	& $\times$1.54	
& 677,511	& $\times$1.11	
& 959,736	& $\times$1.58	
& 713,304	& $\times$1.17	
& 701,901	& $\times$1.15	
& 704,330	& $\times$1.16	
& 692,719	& $\times$1.14	
& 608,402 \\

\midrule
15        & \textbf{smallvec}
& 1,432	& $\times$7.12	
& 1,387	& $\times$6.90	
& 1,514	& $\times$7.53	
& 478	& $\times$2.38	
& 757	& $\times$3.77	
& 659	& $\times$3.28	
& 738	& $\times$3.67	
& 964	& $\times$4.80	
& 2,077	& $\times$10.33	
& 1,804	& $\times$8.98	
& 1,467	& $\times$7.30	
& 776	& $\times$3.86	
& 713	& $\times$3.55	
& 201 \\

\midrule
16        & \textbf{smallvec}
& 1,224	& $\times$3.25	
& 1,688	& $\times$4.48	
& 1,797	& $\times$4.77	
& 1,706	& $\times$4.53	
& 1,513	& $\times$4.01	
& 948	& $\times$2.51	
& 1,474	& $\times$3.91	
& 3,638	& $\times$9.65	
& 2,247	& $\times$5.96	
& 2,306	& $\times$6.12	
& 2,161	& $\times$5.73	
& 1,172	& $\times$3.11	
& 1,193	& $\times$3.16	
& 377 \\

\midrule
17        & \textbf{simple-slab}
& 2,944	& $\times$6.94	
& 1,686	& $\times$3.98	
& 3,395	& $\times$8.01	
& 3,448	& $\times$8.13	
& 2,497	& $\times$5.89	
& 859	& $\times$2.03	
& 886	& $\times$2.09	
& 1,972	& $\times$4.65	
& 4,607	& $\times$10.87	
& 2,298	& $\times$5.42	
& 1,866	& $\times$4.40	
& 1,221	& $\times$2.88	
& 1,670	& $\times$3.94	
& 424 \\

\midrule
18        & \textbf{simple-slab}
& 2,692	& $\times$5.98	
& 2,953	& $\times$6.56	
& 2,800	& $\times$6.22	
& 1,032	& $\times$2.29	
& 2,495	& $\times$5.54	
& 1,492	& $\times$3.32	
& 1,751	& $\times$3.89	
& 1,735	& $\times$3.86	
& 3,606	& $\times$8.01	
& 3,326	& $\times$7.39	
& 2,815	& $\times$6.26	
& 1,847	& $\times$4.10	
& 1,806	& $\times$4.01	
& 450 \\

\midrule
19        & \textbf{ordnung}
& 358,951	& $\times$9.30	
& 103,449	& $\times$2.68	
& 140,733	& $\times$3.64	
& 317,213	& $\times$8.21	
& 253,858	& $\times$6.57	
& 47,245	& $\times$1.22	
& 300,551	& $\times$7.78	
& 138,779	& $\times$3.59	
& 70,302	& $\times$1.82	
& 108,954	& $\times$2.82	
& 105,142	& $\times$2.72	
& 59,525	& $\times$1.54	
& 60,415	& $\times$1.56	
& 38,614 \\

\midrule
20        & \textbf{ordnung}
& 299,830	& $\times$6.72	
& 95,966	& $\times$2.15	
& 97,916	& $\times$2.20	
& 683,809	& $\times$15.34	
& 538,915	& $\times$12.09	
& 60,453	& $\times$1.36	
& 60,918	& $\times$1.37	
& 67,916	& $\times$1.52	
& 61,581	& $\times$1.38	
& 67,771	& $\times$1.52	
& 56,134	& $\times$1.26	
& 53,738	& $\times$1.21	
& 53,301	& $\times$1.20	
& 44,585 \\

\midrule
21        & \textbf{cbox}
& 10,069	& $\times$2.33	
& 14,661	& $\times$3.40	
& 12,414	& $\times$2.88	
& 100,351	& $\times$23.26	
& 7,791	    & $\times$1.81	
& 10,134	& $\times$2.35	
& 9,908	    & $\times$2.30	
& 12,414	& $\times$2.88	
& 8,645	& $\times$2.00	
& 6,313	& $\times$1.46	
& 6,327	& $\times$1.47	
& 4,981	& $\times$1.15	
& 4,999	& $\times$1.16	
& 4,315 \\

\midrule
22        & \textbf{qwutils}
& 1,023,155	& $\times$1.69	
& 708,732	& $\times$1.17	
& 640,438	& $\times$1.06	
& 741,661	& $\times$1.23	
& 660,311	& $\times$1.09	
& 673,418	& $\times$1.12	
& 608,279	& $\times$1.01	
& 691,574	& $\times$1.15	
& 696,784	& $\times$1.15	
& 775,310	& $\times$1.28	
& 687,653	& $\times$1.14	
& 603,341	& $\times$1.00	
& 604,278	& $\times$1.00	
& 603,904 \\

\midrule
23        & \textbf{endian\_trait}
& 400,495	& $\times$6.20	
& 203,874	& $\times$3.16	
& 491,300	& $\times$7.61	
& 127,354	& $\times$1.97	
& 165,922	& $\times$2.57	
& 239,877	& $\times$3.72	
& 601,076	& $\times$9.31	
& 94,597	& $\times$1.47	
& 82,334	& $\times$1.28	
& 157,188	& $\times$2.43	
& 137,761	& $\times$2.13	
& 69,490	& $\times$1.08	
& 68,902	& $\times$1.07	
& 64,569 \\

\midrule
24        & \textbf{pnet\_packet}
& 14,880	& $\times$4.08	
& 13,308	& $\times$3.65	
& 69,505	& $\times$19.06	
& 94,849	& $\times$26.01	
& 29,485	& $\times$8.09	
& 8,536	    & $\times$2.34	
& 10,434	& $\times$2.86	
& 13,780	& $\times$3.78	
& 22,440	& $\times$6.15	
& 22,658	& $\times$6.21	
& 23,124	& $\times$6.34	
& 6,527	    & $\times$1.79	
& 6,426	    & $\times$1.76	
& 3,646 \\

\midrule
25        & \textbf{rdiff}
& 12,640	& $\times$8.88	
& 8,714	    & $\times$6.12	
& 3,165	    & $\times$2.22	
& 15,616	& $\times$10.97	
& 10,208	& $\times$7.17	
& 6,205	    & $\times$4.36	
& 9,828	    & $\times$6.91	
& 4,753 	& $\times$3.34	
& 8,223	& $\times$5.78	
& 8,557	& $\times$6.01	
& 8,135	& $\times$5.72	
& 3,204	& $\times$2.25	
& 3,301	& $\times$2.32	
& 1,423 \\

\midrule
26        & \textbf{through}
& 26,325	& $\times$5.60	
& 12,049	& $\times$2.56	
& 42,333	& $\times$9.01	
& 11,703	& $\times$2.49	
& 17,221	& $\times$3.66	
& 9,938	    & $\times$2.11	
& 7,343	    & $\times$1.56	
& 10,639	& $\times$2.26	
& 26,243	& $\times$5.58	
& 28,397	& $\times$6.04	
& 17,599	& $\times$3.74	
& 9,358	& $\times$1.99	
& 9,241	& $\times$1.97	
& 4,700 \\

\midrule\midrule
\multicolumn{2}{c|}{\textbf{Average}} 
& \afldstime          & \afldsrate
& \aflppdstime        & \aflppdsrate            
& \aflgodstime        & \aflgodsrate        
& \windrangerdstime   & \windrangerdsrate   
& \fishfuzzdstime     & \fishfuzzdsrate     
& \aflrundstime       & \aflrundsrate       
& \lysodstime         & \lysodsrate         
& \panickillerdstime  & \panickillerdsrate  
& \rustgodissafedstime & \rustgodissafedsrate 
& \rustgobasedstime    & \rustgobasedsrate
& \rustgooptdstime     & \rustgooptdsrate
& \rustgoprundstime    & \rustgoprundsrate 
& \rustgosyncdstime    & \rustgosyncdsrate
& \rustgodstime         \\
\bottomrule
\end{tabular}%
}
\end{table*}

\subsection{Target Path Pruning}
\label{ss:target-pruning}
We evaluate how target-specific reachability analysis effectively eliminates unreachable paths from the application entry point to each vulnerable target, using the same benchmark as in \Sref{ss:target-optimization}.
As shown in~\Tref{tbl:target-pruning}, \sys prunes \rustgopr of unreachable paths on average (31,509 out of \totalbb), while \rustgobase and \rustgoopt eliminate \rustgobasepr (20,647 out of \totalbb) and \rustgooptpr (21,357 out of \totalbb), respectively.
\sys eliminates \rustgobaseprapproximate more unreachable paths on average, compared to \rustgobase.
As mentioned in~\Sref{sss:fine-grained-ablation}, \rustgobase relies on the existing reachability-based analysis, which is not suitable to Rust programs due to its structurally complex Rust’s standard library calls, often resulting in overly conservative analysis. 
Also, \sys eliminates \rustgooptprapproximate more irrelevant paths to targets on average compared to \rustgoopt, which extends \rustgobase by applying post-dominance-based target consolidation.

\rustgoprun shows the same target-path pruning rate as \sys, since both employ the same target-specific path pruning techniques, as mentioned in~\Tref{tbl:ablation-module}.
In the case of \rustgosync, even when we additional apply the async/await function–aware pruning, it shows the same performance as \rustgoprun and \sys.
This is because functions related to asynchronous or concurrent execution (\ie \cc{async/await}), which are excluded from target-specific path pruning (\Sref{ss:DesignModule2}), account for only 5.90\% of the total functions on average (219 out of 3,712).
These few functions are entirely removed during RustGo’s target optimization phase (\Sref{ss:DesignModule1}), resulting in the same set of targets as \sys and thus producing identical results when target-specific path pruning is applied.

\subsection{Dynamic Fair Toggling}
\label{ss:effectiveness-of-dynamic-toggling}
We evaluate how dynamic fair toggling achieves fair and target-specific directed fuzzing across multiple targets.
Dynamic toggling is the first approach to enable multi-target directed fuzzers to maintain per-target directness while also achieving fair fuzzing across targets. 
To achieve this, not only does it apply a round-robin scheduling mechanism, but it also ensures that each target is fuzzed with preserved directness. 
Specifically, \sys's dynamic toggling incorporates dynamic pruning, which efficiently and accurately identifies target-relevant code for each target and dynamically restricts exploration to only allowed code regions upon switching targets. 
For target fairness, we follow AFLRUN’s evaluation methodology and measured the standard deviation of energy (mutation count) per target. 
Specifically, for each of the 15 targets, we record the number of mutations assigned over a 24-hour and computed the standard deviation of mutation counts across targets to assess whether mutations are distributed fairly among target locations.
As shown in ~\Tref{tbl:dynamic-toggling-evaluation} in the Appendix \Sref{apen:evaluation-result-details}, \sys achieve a lower deviation (1.64) compared to AFLRUN (4.88), indicating more balanced fuzzing per each targets. 
This ensures that each target receives a balanced share of fuzzing efforts without sacrificing directedness. 
%

\subsection{Effectiveness of \sys}
\label{ss:effectiveness-of-rustgo}
We evaluate the vulnerability detection efficiency of \sys by measuring the time it takes to cover injected vulnerabilities, and compare the results against two code-coverage-based fuzzers and six state-of-the-art directed fuzzers. These results are presented in \Tref{tbl:performance}, and in \Fref{fig:performance} in the Appendix~\Sref{apen:evaluation-figures}.
In \Tref{tbl:performance}, \cc{Time} denotes the absolute fuzzing time required to trigger all 15 inserted bugs. 
This time value is calculated by averaging the fuzzing times taken to expose each of the 15 bugs. 
The \cc{Rate} represents the arithmetic mean of the per-bug detection time ratios between a given fuzzer and \sys. 
Specifically, for each inserted bug, we compute the ratio of the time required by the fuzzer to that required by \sys. 
These per-bug ratios are then averaged across all inserted bugs to obtain the final Rate value. \cc{Total Time Ratio (TTR)} represents the ratio of the total detection time of a fuzzer to that of \sys, summed across all inserted bugs.

\renewcommand{\arraystretch}{1.0} 
\begin{table}[t]
  \centering
  \caption{Previously Unknown Bugs Discovered by \sys, with a comparative evaluation against fuzz driver generators including RPG, RUG, and deepSURF.  Driver represents the number of generated drivers, and Detect indicates whether the generated drivers successfully detect the vulnerability, where \textcolor{red}{\ding{55}} denotes failure to detect and \textcolor{green}{\ding{52}} denotes successful detection.}
  \label{tbl:new-vulnerability-fuzz-driver-generator}

\begingroup

\renewcommand{\arraystretch}{1.0}
\setlength{\tabcolsep}{2pt}
\huge

\def\DriverHead{%
  \shortstack[c]{%
    \textbf{Driver}\\
    \textbf{(\#)}%
  }%
}

\def\DetectHead{%
  \shortstack[c]{%
    \textbf{Detect}\\
    \mbox{%
      \textbf{(}%
      \textcolor{red}{\ding{55}}%
      \hspace{0.15em}\textbf{|}\hspace{0.15em}%
      \textcolor{green}{\ding{52}}%
      \textbf{)}%
    }%
  }%
}

\resizebox{\linewidth}{!}{%
\begin{tabular}{
  >{\centering\arraybackslash}m{5.0cm}|
  >{\centering\arraybackslash}m{2.0cm}|
  >{\centering\arraybackslash}m{2.5cm}|
  >{\centering\arraybackslash}m{2.0cm}|
  >{\centering\arraybackslash}m{2.5cm}|
  >{\centering\arraybackslash}m{2.0cm}|
  >{\centering\arraybackslash}m{2.5cm}|
  >{\centering\arraybackslash}m{2.5cm}
}
\toprule

\multirow{3}{*}{\textbf{Program}}
& \multicolumn{6}{c|}{\textbf{Fuzz Driver Generator}}
& \multicolumn{1}{c}{
    \multirow{2}{*}{\textbf{RustGo}}}
\\

\cline{2-7}

& \multicolumn{2}{c|}{\textbf{RPG}}
& \multicolumn{2}{c|}{\textbf{RUG}}
& \multicolumn{2}{c|}{\textbf{deepSURF}}
\\

\cline{2-8}

& \DriverHead
& \DetectHead
& \DriverHead
& \DetectHead
& \DriverHead
& \DetectHead
& \DetectHead
\\

\midrule\midrule

\textbf{ouch}
& 3
& \textcolor{red}{\ding{55}}
& 35
& \textcolor{red}{\ding{55}}
& 0
& \textcolor{red}{\ding{55}}
& \textcolor{green}{\ding{52}}
\\
\midrule

\textbf{grcov}
& 21
& \textcolor{red}{\ding{55}}
& 2
& \textcolor{red}{\ding{55}}
& 0
& \textcolor{red}{\ding{55}}
& \textcolor{green}{\ding{52}}
\\
\midrule

\textbf{fast-float-2}
& 14
& \textcolor{green}{\ding{52}}
& 45
& \textcolor{green}{\ding{52}}
& 0
& \textcolor{red}{\ding{55}}
& \textcolor{green}{\ding{52}}
\\
\midrule

\textbf{fast-float}
& 14
& \textcolor{green}{\ding{52}}
& 45
& \textcolor{green}{\ding{52}}
& 0
& \textcolor{red}{\ding{55}}
& \textcolor{green}{\ding{52}}
\\
\midrule

\textbf{dbn}
& 12
& \textcolor{red}{\ding{55}}
& -
& -
& 0
& \textcolor{red}{\ding{55}}
& \textcolor{green}{\ding{52}}
\\
\midrule

\textbf{xams-elf (get-bucket)}
& 66
& \textcolor{red}{\ding{55}}
& 2
& \textcolor{red}{\ding{55}}
& 1
& \textcolor{red}{\ding{55}}
& \textcolor{green}{\ding{52}}
\\
\midrule

\textbf{xams-elf (get-chain)}
& 66
& \textcolor{red}{\ding{55}}
& 2
& \textcolor{red}{\ding{55}}
& 1
& \textcolor{red}{\ding{55}}
& \textcolor{green}{\ding{52}}
\\
\midrule

\textbf{pilka}
& 1
& \textcolor{red}{\ding{55}}
& 21
& \textcolor{green}{\ding{52}}
& 0
& \textcolor{red}{\ding{55}}
& \textcolor{green}{\ding{52}}
\\
\midrule

\textbf{aarty}
& 22
& \textcolor{green}{\ding{52}}
& 2
& \textcolor{red}{\ding{55}}
& 0
& \textcolor{red}{\ding{55}}
& \textcolor{green}{\ding{52}}
\\
\midrule

\textbf{kalker}
& 33
& \textcolor{red}{\ding{55}}
& 0
& \textcolor{red}{\ding{55}}
& 0
& \textcolor{red}{\ding{55}}
& \textcolor{green}{\ding{52}}
\\
\midrule

\textbf{ethaddregen}
& 1
& \textcolor{red}{\ding{55}}
& 7
& \textcolor{red}{\ding{55}}
& 0
& \textcolor{red}{\ding{55}}
& \textcolor{green}{\ding{52}}
\\
\midrule

\textbf{infisearch (common)}
& 0
& \textcolor{red}{\ding{55}}
& 38
& \textcolor{red}{\ding{55}}
& 8
& \textcolor{red}{\ding{55}}
& \textcolor{green}{\ding{52}}
\\
\midrule

\textbf{infisearch (search)}
& 0
& \textcolor{red}{\ding{55}}
& 38
& \textcolor{red}{\ding{55}}
& 8
& \textcolor{red}{\ding{55}}
& \textcolor{green}{\ding{52}}
\\

\midrule\midrule

\textbf{Average}
& 19.46
& 3 / 13
& 19.75
& 3 / 12
& 1.38
& 0 / 13
& 13 / 13
\\

\bottomrule
\end{tabular}%
}

\endgroup
\end{table}

\sys shows the best performance among all eight compared fuzzers, demonstrating an average \totalaveragespeedup speedup in vulnerability exposure across the \benchnum evaluation applications.
Specifically, \sys is \afl and \aflpp faster than the coverage-guided fuzzers AFL and AFL++, respectively. Compared to directed fuzzers, \sys shows average speedups of \aflgo, \windranger, \fishfuzz, \aflrun, \lyso, and \panickiller over AFLGo, Windranger, FishFuzz, AFLRUN, Lyso, and PanicKiller, respectively.
In particular, \sys shows a maximum $\times$ 10.16 speedup on the \cc{gimli} application. This improvement is mainly attributed to {\sys}’s post-dominant target consolidation, which merged 15 injected locations into a single semantically representative target, and its target-specific analysis that eliminated 80.43\% of unreachable blocks (\ie pruning 40,377 out of 50,204 basic blocks), thereby enabling fuzzing energy to be concentrated more effectively. 

Furthermore, even when compared with Lyso and AFLRUN, a state-of-the-art directed fuzzer designed for multi-target scenarios, \sys achieves average speedups of \lyso and \aflrun, with maximum speedups of $\times$ 4.25 and $\times$ 4.67, respectively, validating the efficiency of our proposed techniques.
These results indicate that target path pruning and dynamic toggling of \sys provide more efficient guidance for exploring multiple target locations in \rust programs, compared to path-diversity metric and unbiased energy assignment of AFLRUN and multi-step, multi-target guidance of Lyso.

In comparison with PanicKiller, \sys achieves an average speedup of $\times$ 2.09 (up to $\times$ 3.75), indicating more accurate exploration of vulnerable paths in Rust programs.
PanicKiller employs the target selection analysis proposed in prior work~\cite{min2024erasan}, which suffers from over-approximation due to context-insensitive analysis of standard library. As a result, across the 13 benchmark programs, PanicKiller identifies 2,585 out of 3,664 functions (70.55\%) as fuzzing targets, causing the fuzzer to explore irrelevant paths.
\sys also shows an average performance improvement of \afl (up to $\times$ 9.34) compared to AFL. When compared to AFL++, the only modern coverage-based fuzzer that can be directly applied to Rust applications, \sys shows a speedup of \aflpp (up to $\times$ 5.45).
These results demonstrate that maximizing code coverage without considering Rust-specific memory safety characteristics can lead to inefficient fuzzing by wasting energy on memory-safe areas.

Additionally, as shown in ~\Tref{tbl:deepsurf-benchmark-total}, \sys demonstrates improved vulnerability detection speed compared to AFL, AFL++, AFLGo, WindRanger, FishFuzz, AFLRUN, Lyso, and PanicKiller on the 26 deepSURF benchmarks, with performance gains of \afldsrate, \aflppdsrate, \aflgodsrate, \windrangerdsrate, \fishfuzzdsrate, \aflrundsrate, \lysodsrate, and \panickillerdsrate, respectively.
In addition, \sys achieves an average pruning rate of 88.75\% across the 26 programs, eliminating 3,875 out of 4,366 basic blocks. 
\sys achieves a maximum speedup of $\times$ 26.01 over existing directed fuzzer (\ie WindRanger), with the largest performance improvement observed on bug\#24-\cc{pnet\_packet} (\ie a 96.16\% reduction).
These results demonstrate that \sys consistently outperforms prior fuzzing approaches on real-world Rust vulnerabilities.

\renewcommand{\arraystretch}{1.0} 
\begin{table*}[t]
  \centering
  \caption{Unknown bugs discovered by \sys. TTE (ms) represents the time-to-exposure of a fuzzer to that of \sys. Rate represents detection time ratio of a fuzzer to those of \sys. } 
  \label{tbl:new-vulnerability-evaluation-2}
  \begingroup
\setlength{\tabcolsep}{2pt}   
\renewcommand{\arraystretch}{1.0}
\huge
\resizebox{\textwidth}{!}{%
\begin{tabular}{
>{\centering\arraybackslash}m{5.0cm}| 
>{\centering\arraybackslash}m{4.0cm}|
>{\centering\arraybackslash}m{2.4cm}|
>{\centering\arraybackslash}m{4.8cm}|
>{\centering\arraybackslash}m{2.4cm}|
>{\centering\arraybackslash}m{1.8cm}|
>{\centering\arraybackslash}m{2.4cm}|
>{\centering\arraybackslash}m{1.8cm}|
>{\centering\arraybackslash}m{2.4cm}|
>{\centering\arraybackslash}m{1.8cm}|
>{\centering\arraybackslash}m{2.4cm}|
>{\centering\arraybackslash}m{1.8cm}|
>{\centering\arraybackslash}m{2.4cm}|
>{\centering\arraybackslash}m{1.8cm}|
>{\centering\arraybackslash}m{2.4cm}|
>{\centering\arraybackslash}m{1.8cm}|
>{\centering\arraybackslash}m{2.4cm}|
>{\centering\arraybackslash}m{1.8cm}|
>{\centering\arraybackslash}m{2.4cm}|
>{\centering\arraybackslash}m{1.8cm}|
>{\centering\arraybackslash}m{2.4cm}|
>{\centering\arraybackslash}m{1.8cm}
}
\toprule
\multirow{3}{*}{\textbf{Program}} 
& \multicolumn{3}{c|}{\textbf{Info}} 
& \multicolumn{6}{c|}{\textbf{Compared Fuzzer Approaches}} 
& \multicolumn{10}{c|}{\textbf{Variants for ablation study}} 
& \multicolumn{2}{c}{\multirow{2}{*}{\textbf{RustGo}} }  \\
\cline{2-20}

& \multirow{2}{*}{\textbf{Root Cause}} 
& \multirow{2}{*}{\textbf{Status}} 
& \multirow{2}{*}{\textbf{ID}} 
& \multicolumn{2}{c|}{\textbf{AFL++}} 
& \multicolumn{2}{c|}{\textbf{AFLRUN}} 
& \multicolumn{2}{c|}{\textbf{PanicKiller}} 
& \multicolumn{2}{c|}{\textbf{\rustgodissafe}} 
& \multicolumn{2}{c|}{\textbf{\rustgobase}} 
& \multicolumn{2}{c|}{\textbf{\rustgoopt}} 
& \multicolumn{2}{c|}{\textbf{\rustgoprun}} 
& \multicolumn{2}{c|}{\textbf{\rustgosync}} \\
\cline{5-22}
&  &  &  
& \textbf{TTE (ms)} & \textbf{Rate ($\times$)}
& \textbf{TTE (ms)} & \textbf{Rate ($\times$)}
& \textbf{TTE (ms)} & \textbf{Rate ($\times$)}
& \textbf{TTE (ms)} & \textbf{Rate ($\times$)}
& \textbf{TTE (ms)} & \textbf{Rate ($\times$)}
& \textbf{TTE (ms)} & \textbf{Rate ($\times$)}
& \textbf{TTE (ms)} & \textbf{Rate ($\times$)}
& \textbf{TTE (ms)} & \textbf{Rate ($\times$)}
& \textbf{TTE (ms)} & \textbf{Rate ($\times$)} \\
\midrule\midrule

\textbf{ouch} & Read to Uninit Memory & Patched & \makecell[c]{RUSTSEC-2024-0374 \\ (CVE-2024-13941)} 
& 2,784,816 & 3.03
& 1,843,953 & 2.01
& 1,384,144 & 1.51
& 2,461,778	& 2.68	
& 2,068,186	& 2.25	
& 1,951,771	& 2.13	
& 1,316,778	& 1.43	
& 1,316,778	& 1.43	
& 918,475 & 1.00 \\
\midrule

\textbf{grcov} 
& Heap BoF & Patched & RUSTSEC-2025-0005 
& TO & - 
& 3,829,007 & 1.54
& 45,176,157 & 18.13
& TO	& -	
& TO	& -	
& TO	& -	
& 47,714,080 & 19.15	
& 47,714,080 & 19.15
& 2,491,889 & 1.00 \\
\midrule

\textbf{fast-float-2} 
& Null Pointer Deref & Patched & RUSTSEC-2025-0002
& 24,325 & 10.87	
& 23,139 & 10.34
& 15,889 & 7.10
& 24,183 & 10.81
& 14,223 & 6.36
& 11,687 & 5.22
& 8,962 & 4.00
& 8,962 & 4.00
& 2,238	& 1.00 \\
\midrule

\textbf{fast-float} 
& Null Pointer Deref & Patched & RUSTSEC-2025-0003
& 24,325 & 10.87	
& 23,139 & 10.34
& 15,889 & 7.10
& 24,183 & 10.81
& 14,223 & 6.36
& 11,687 & 5.22
& 8,962 & 4.00
& 8,962 & 4.00
& 2,238	& 1.00 \\
\midrule

\textbf{dbn} 
& Heap BOF & Patched & RUSTSEC-2024-0377
& 2,124,689 & 4.96	
& 915,029 & 2.14
& 1,354,593 & 3.16
& 1,600,405 & 3.74
& 1,219,104 & 2.85
& 995,417 & 2.32
& 825,782 & 1.93
& 825,782 & 1.93
& 428,192	& 1.00 \\
\midrule

\textbf{xams-elf (get-bucket)} 
& Stack BOF & Patched & RUSTSEC-2025-0018
& 353,812 & 1.33
& 1,689,570 & 6.37
& 1,619,558 & 6.10
& 1,185,108 & 4.46
& 1,017,860 & 3.83
& 964,428 & 3.63
& 434,693 & 1.64
& 434,693 & 1.64
& 265,425 & 1.00 \\
\midrule

\textbf{xams-elf (get-chain)} 
& Stack BOF & Patched & RUSTSEC-2025-0018
& 345,733 & 1.37
& 766,093 & 3.03	
& 889,614 & 3.51
& 844,519 & 3.34
& 872,979 & 3.45
& 863,256 & 3.41
& 476,145 & 1.88
& 476,145 & 1.88
& 253,099 & 1.00 \\
\midrule

\textbf{pilka} 
& Null Pointer Deref & Patched & - 
& 37,820	& 3.62	
& 33,896	& 3.24	
& 43,180	& 4.13
& 42,635 & 4.08
& 64,711 & 6.19
& 50,246 & 4.80
& 49,864 & 4.77
& 49,864 & 4.77
& 10,459	& 1.00 \\
\midrule

\textbf{aarty} 
& Heap BOF & Confirmed & - 
& 308,347	& 1.52	
& 259,771	& 1.28	
& 240,468	& 1.18	
& 315,178 & 1.55
& 304,437 & 1.50
& 278,593 & 1.37
& 272,417 & 1.34
& 272,417 & 1.34
& 203,127  & 1.00 \\
\midrule

\textbf{kalker} 
& Stack UAF & Confirmed & - 
& 454,147	& 9.85	
& 98,590	& 2.14	
& 46,880	& 1.02
& 281,076 & 6.09
& 208,910 & 4.53
& 201,767 & 4.37
& 56,123 & 1.22
& 56,123 & 1.22
& 46,125  & 1.00 \\
\midrule

\textbf{ethaddregen} 
& Null Pointer Deref & Reported & - 
&  2,038,816	& 4.93	
& 734,318	& 1.78	
& 1,549,075	& 3.75
& 704,363 & 1.70
& 2,024,454 & 4.90
& 2,066,430 & 5.00
& 859,888 & 2.08
& 859,888 & 2.08
& 413,553 & 1.00 \\
\midrule

\textbf{infisearch (common)}
& Heap BOF & Reported & - 
& 412,406	& 2.50	
& 463,825	& 2.81	
& 400,875	& 2.43	
& 331,678 & 2.01
& 205,889 & 1.25
& 206,425 & 1.25
& 206,599 & 1.25
& 206,599 & 1.25
& 165,293 & 1.00 \\
\midrule

\textbf{infisearch (search)}
& Global BOF & Reported & - 
& 193,237	& 1.79	
& 147,950	& 1.37	
& 477,016	& 4.43	
& 569,466 & 5.29
& 261,678 & 2.43
& 219,940 & 2.04
& 217,609 & 2.02
& 217,609 & 2.02
& 107,713	& 1.00 \\
\midrule\midrule

\textbf{Average} 
& -  & -  & - 
& \aflppnewvultime	& \aflppnewvul
& \aflrunnewvultime	& \aflrunnewvul
& \panickillernewvultime	& \panickillernewvul
& \rustgodissafenewvultime	& \rustgodissafenewvul
& \rustgobasenewvultime  & \rustgobasenewvul
& \rustgooptnewvultime   & \rustgooptnewvul
& \rustgoprunnewvultime  & \rustgoprunnewvul
& \rustgosyncnewvultime  & \rustgosyncnewvul
& \rustgonewvultime	& 1.00 \\
\bottomrule
\end{tabular}%
}
\endgroup

\end{table*}

\subsection{Ablation Study}
\label{ss:ablation-study}
We conduct an ablation study to understand how each component of \sys contributes to its overall performance. As detailed in~\Sref{sss:fine-grained-ablation}, we prepare six variants as follows: \rustgodissafe, \rustgobase, \rustgoopt, \rustgoprun, \rustgosync, and \sys which incrementally enable one of the proposed components.
As shown in \Tref{tbl:ablation}, and in \Fref{fig:ablation} in the Appendix~\Sref{apen:evaluation-figures} and the \cc{httparse} case study (\Tref{tbl:ablation}) in the Appendix~\Sref{apen:httparse-case-study}, \rustgodissafe, \rustgobase, \rustgoopt, \rustgoprun, and \rustgosync take \rustgodissaferesult, \rustgobaseresult, \rustgooptresult, \rustgoprunresult, and \rustgosyncresult longer than \sys, respectively, clearly indicating that \sys achieves the most efficient fuzzing performance through the full integration of its design.
Of the total performance improvement $\times$ 2.07 achieved by \sys , target optimization (\Sref{ss:DesignModule1}), target-specific path pruning (\Sref{ss:DesignModule2}), and dynamic fair toggling fuzzing (\Sref{ss:DesignModule3}) contribute approximately 22.70\% ($\times$ 0.47), 59.90\% ($\times$ 1.24), and 17.39\% ($\times$ 0.36), respectively.

From the benchmark programs, \rustgodissafe identifies 62.81\% of the code as safe on average, collecting partial coverage feedback only from the remaining 37.18\% unsafe code (1,380 out of 3,712).
Compared to \sys, the bug detection speed dropped by \rustgodissafe, and even against native AFL++, the performance decreased by $\times$ 0.42. 
These results suggest that excluding coverage instrumentation for safe code, which constitutes the majority of Rust programs, leads to a significant loss in useful feedback, and thus reduced fuzzing effectiveness. Moreover, since UAF bugs can also occur in safe Rust code, omitting coverage feedback for these regions may introduce additional false negatives. 
Also, \rustgobase and \rustgoopt take \rustgobaseresult and \rustgooptresult longer than \sys, respectively. \rustgobase results in a \rustgobaseresult runtime overhead, achieving only a \rustgobasepr pruning rate across our benchmark programs, whereas \sys attains \rustgopr. Moreover, \rustgoopt leads to a $\times$ 0.47 performance improvement than \rustgobase in eliminating redundant targets.

Additionally, in multi-threaded applications, \rust's async/\allowbreak await constructs introduce nontrivial lifetime interactions that can affect the identification of temporal memory safety bug targets. To overcome this, \rustgosync pre-identifies functions involving asynchronous or concurrent execution and disables pruning for code involving raw pointers or aliasing relationships within them, to avoid false negatives.
Across our benchmark programs, the \rustgosync identifies an average of 219 \cc{async/await} functions (out of 3,664).
However, due to \sys's target optimization, the number of target locations initially increased by \cc{async/await} handling converges in most cases to the same final count as in \rustgoprun (\ie concurrency-safe mode disabled), demonstrating that \sys maintains comparable runtime performance even with concurrency-safe mode enabled.

Additionally, as shown in~\Tref{tbl:deepsurf-benchmark-total}, our ablation study on the 26 deepSURF benchmarks shows improvements of \rustgodissafedsrate, \rustgobasedsrate, \rustgooptdsrate, \rustgoprundsrate, and \rustgosyncdsrate over \rustgodissafe, \rustgobase, \rustgoopt, \rustgoprun, and \rustgosync, respectively.
The three main components of \sys contribute performance improvements of 16.60\% ($\times$ 0.43), 44.02\% ($\times$ 1.14), and 39.38\% ($\times$ 1.02), respectively, consistent with the contribution pattern observed in our constructed benchmark.
These results indicate that each component of \sys remains effective on real-world Rust programs, and that the fully integrated variants are necessary to achieve maximum fuzzing performance.

\subsection{Comparison with Fuzz Driver Generator}
\label{ss:comparison-with-fuzz-target-generator}
We evaluate \rust fuzz driver generators on 13 newly discovered bugs identified by \sys. Among the eight existing \rust fuzz driver generators~\cite{xu2024rpg,jiang2021rulf,syrust,crabtree,fries,rumono,rug,deepsurf}, we select three publicly available fuzz driver generators, including deepSURF~\cite{deepsurf} (state-of-the-art approach), RPG~\cite{xu2024rpg}, and RUG~\cite{rug}.
Their effectiveness is measured by whether the generated drivers can trigger these new memory safety bugs. Since fuzz driver generators are designed to produce fuzzing drivers rather than to schedule inputs or allocate fuzzing resources across multiple drivers, optimizing fuzzing efficiency using multiple generated drivers is beyond the scope of this work. 
Accordingly, in our comparison with fuzz driver generators, we evaluate only whether each generated driver can detect these bugs within a 24-hour time budget.
As shown in \Tref{tbl:new-vulnerability-fuzz-driver-generator}, although fuzz driver generators generate an average of 13 fuzz drivers per program, they detect only 15.38\% (\ie 2 out of 13) of the memory violation vulnerabilities. 
As detailed in Appendix \Sref{apen:comparison-with-fuzz-driver-generator}, this limitation arises mainly because existing tools cannot construct complex argument types required to trigger real-world memory violations (\eg \cc{BTreeMap}), nor can they precisely capture the vulnerability-triggering execution context (\eg UAF bugs).

\subsection{New Vulnerability Detection}
\label{ss:new-vulnerability-discovery}
We conduct evaluations on diverse Rust applications collected from the official Rust package registry, crates.io~\cite{web:crateio}, to assess \sys's effectiveness in real-world memory safety bug discovery.
Our evaluation targets include both executable binaries and library-style crates. 
As summarized in \Tref{tbl:new-vulnerability-evaluation-2}, \sys successfully uncovered 13 previously unknown memory bugs across real-world Rust applications. 
We have responsibly reported all discovered vulnerabilities to the corresponding vendors. Among them, currently, ten bugs have been confirmed, with eight already patched upstream.
The remaining three are currently undergoing coordination with developers. Of the confirmed issues, \rustsecnum have been assigned RustSec advisories, and one has been issued a CVE ID.

We conduct additional comparisons with other fuzzers using these newly discovered bugs.
Since the newly discovered vulnerabilities can only be reproduced on Rust programs requiring LLVM 14 or newer, many DGFs (\eg AFLGo, Windranger, FishFuzz, and  Lyso) cannot be directly compared due to their lack of LLVM 14 support. We therefore compare \sys against AFL++, AFLRUN, and PanicKiller, which fully support LLVM 14.
As showed as in \Tref{tbl:new-vulnerability-evaluation-2}, \sys outperforms AFL++, AFLRUN, and PanicKiller in detecting the 13 newly discovered vulnerabilities, achieving average speedups of \aflppnewvul, \aflrunnewvul, and \panickillernewvul, respectively.

To further evaluate  the contribution of variants in \sys, we extend an ablation study on these  newly discovered bugs. 
\sys consistently outperforms its ablated variants, including \rustgodissafe, \rustgobase, \rustgoopt, \rustgoprun, and \rustgosync, achieving average \allowbreak speedups of \rustgodissafenewvul, \rustgobasenewvul, \rustgooptnewvul, \rustgoprunnewvul, and \rustgosyncnewvul, respectively, in detecting the newly discovered vulnerabilities.
Notably, \rustgodissafe, \rustgobase, and \rustgoopt fail to detect the \cc{grcov} vulnerability within the 24-hour time, highlighting the necessity of integrating all modules in \sys.

\section{Discussion}

\noindent \textbf{Fair Multi-target DGF.}
\sys proposes a fair multi-target DGF framework tailored to \rust.
However, distributing fuzzing resources equally across all targets including unreachable vulnerable code is not ideal.
To address this, \sys performs reachability analysis to exclude unreachable targets.
It is also important to consider cases where targets are technically reachable but hard to reach due to complex constraints.
\sys, by default, provides equal fuzzing effort to such targets, assuming that they are difficult but still potentially reachable.
However, depending on the fuzzing purpose, testers can also modify their fuzzing policy based on their criteria instead of prioritizing fairness.

\noindent \textbf{\rust's Memory Safety Guarantees.} 
\rust is a memory-safe language, but critical memory bugs can still occur in unsafe-related regions. Fortunately, these vulnerabilities are limited to a small portion of \rust code typically less than 10\%. As a result, unlike C/C++, where the entire codebase requires testing, \rust enables targeted testing of only a subset of code, significantly reducing testing resources and time. This still can make \rust one of the most effective solutions for addressing memory bugs.

\noindent \textbf{False-Negatives and False-Positives.}
\sys is designed to detect both spatial and temporal memory safety bugs. As mentioned in existing works, such as ERASan~\cite{min2024erasan}, spatial memory safety bugs can only occur through raw pointer dereferences. \sys conservatively identifies all such raw pointer dereference sites at the MIR level and always includes them as targets (even after pruning), ensuring no false negatives. For temporal bugs, which can involve safe references aliased with raw pointers, we apply Andersen’s points-to analysis in a may-alias, context-, field-, and flow-insensitive manner. This conservative analysis is theoretically free from false negatives. While false negatives may still occur due to its implementation limitations, we did not observe any such cases during our evaluation.

\noindent \textbf{Other Bug Types supported by \sys.} 
\sys is designed to detect memory bugs in Rust. However, it can also identify bugs such as type confusion or logic bugs when they result in invalid memory accesses. In addition, the Rust-specific static analysis techniques introduced in RustGo (\eg call-site remapping in \Sref{ss:target-classificatoin-remapping}) and the dynamic fair toggling approaches (\Sref{ss:DesignModule3}) can be applied to detect other types of bugs, such as race conditions and information leaks.

\noindent \textbf{Differences from Fuzz Driver Generators.}
\sys fundamentally differs from fuzz driver generators, as it is designed as a whole-program fuzzer (\eg AFL). Existing fuzz driver generators (\eg libFuzzer) need to prepare appropriate API call sequences and valid argument values, including complicated types. In contrast, \sys is a whole-program fuzzer that starts execution from the program entry point (\ie main) rather than a specific API. Consequently, \sys can naturally initialize valid arguments and explore all possible API call sequences starting from main.

\section{Related Work}

\subsection{Memory Bug Detection}

\noindent \textbf{Static Analysis.}
Several approaches have performed static analysis to detect potential bugs in Rust programs~\cite{bae2021rudra, cui2021safedrop, li2021mirchecker, hua2021rupair}. 
However, unlike \sys, these static analysis-based approaches have difficulty detecting complex memory bugs (\eg use-after-free) and frequently generate many false positive issues similar to those in C/C++.

\noindent \textbf{Dynamic Analysis.} 
Dynamic analysis-based approaches are widely used for memory bug detection, generally due to their higher accuracy.
Miri~\cite{web:miri} and MiriLLI~\cite{mirilli} interpret Rust’s MIR to detect undefined behaviors such as use-after-free and pointer aliasing violations. However, its interpreter-based execution introduces significant runtime overhead and limits its applicability to real-world applications due to incomplete support for multi-threading and external system interactions.
ERASan~\cite{min2024erasan} and RustSan~\cite{cho2024rustsan} customize Address Sanitizer~\cite{serebryany2012addresssanitizer}, the most widely used memory detection approach, for Rust applications by selectively instrumenting programs to reduce redundant and unnecessary memory access checks.
However, while these approaches offer high accuracy, they still suffer from the traditional limitation of dynamic analysis---low coverage---due to the need to secure sufficient workloads.
This issue can be alleviated by integrating them with \sys, the new Rust-specific fuzzing framework.

\subsection{Fuzzing for Rust}

\noindent \textbf{Coverage-guided Fuzzing.}
Few coverage-guided greybox fuzzing approaches have been proposed for \rust.
afl.rs~\cite{web:afl.rs} offers Rust bindings that enable the integration of the widely used greybox fuzzer AFL++ into \rust applications.
cargo-fuzz~\cite{web:cargo-fuzz} enables fuzz testing of Rust libraries using libFuzzer~\cite{web:libfuzzer} as its backend.
honggfuzz-rs~\cite{web:hong-fuzz} offers the necessary interfaces for Rust to interact with Honggfuzz~\cite{web:hongfuzz-orig}.

\noindent \textbf{Directed Greybox Fuzzing.}
Directed Greybox Fuzzing (DGF) approaches~\cite{bohme2017directed, chen2018hawkeye, lee2021constraint, shah2022mc2, du2022windranger, huang2022beacon, srivastava2022one, zong2020fuzzguard, luo2023selectfuzz, kim2023dafl, xiang2024critical, osterlund2020parmesan, huang2024titan, zheng2023fishfuzz, rong2024toward, panickiller} aim to guide fuzzing efforts toward specific target locations, rather than maximizing overall coverage. 
These DGF approaches are originally designed for C/C++, and applying them to Rust environments introduces various challenges, such as automated target selection, precise path pruning, and balanced directed fuzzing. \sys overcomes these challenges by introducing the new framework that automatically identifies vulnerable targets and fuzzes these targets efficiently and fairly. 
Also, although PanicKiller~\cite{panickiller} aims to bypass panic checks while focusing on unsafe code regions, it fails to analyze the precise target locations where memory bugs can actually be triggered, leading to wasted effort on non-buggy unsafe regions and missed memory safety bugs that may be triggered in safe code.

\noindent \textbf{Fuzz Driver Generation.}
Several driver-generation approaches~\cite{jiang2021rulf,syrust,crabtree,fries, rumono, xu2024rpg, rug, deepsurf} have been proposed to automatically generate fuzz drivers for Rust libraries. 
However, existing fuzzing approaches primarily focus on maximizing code coverage without accounting for Rust’s unique characteristics, resulting in wasted effort on safe code, an inability to construct complex argument types required to trigger real-world memory violations, or failure to consider complicated vulnerability-triggering execution contexts (\eg UAF bugs).
In contrast, \sys proposes a Rust-specific fuzzing approach that focuses on vulnerable code regions, significantly reducing unnecessary fuzzing overhead while effectively handling complex types by obtaining type information at runtime and preserving all vulnerability-triggering execution paths.

\section{Conclusion}

Rust has emerged as a fundamental solution to eliminate memory bugs.
However, its guarantees of memory safety apply only to safe code regions. Memory bugs can still occur in unsafe-related regions, which, although small, are security-critical.
Without considering this Rust's unique characteristic, existing fuzzers treat all code the same, wasting fuzzing time on parts that are already memory safe.

This paper presents \sys, the new Rust-directed greybox fuzzer that uses Rust-specific static analysis to identify and merge targets where memory bugs may occur and prune irrelevant paths, and introduces a new fuzzing framework to focus on targets fairly prone to memory bugs.
Our evaluation shows that \sys prunes \rustgopr of irrelevant paths, reaches targets \panickiller to \windranger faster than existing fuzzers, and identifies 13 previously unknown bugs (\rustsecnum RUSTSEC-assigned, one CVE-assigned).  
These results highlight \sys's effectiveness in improving both precision and efficiency in the context of \rust fuzzing.

\begin{acks}
This work was supported by the National Research Foundation of Korea (NRF) grants funded by the Korea government (MSIT) (No. RS-2025-24535159, RS-2026-25497375) and the Institute of Information \& Communications Technology Planning \& Evaluation (IITP) grants funded by the Korea government (MSIT) (No. RS-2024-00437306, RS-2024-00341722, RS-2024-00439819, RS-2024-00337414). This work was also supported by the IITP-ITRC (Information Technology Research Center) grant funded by the Korea government (MSIT) (IITP-2026-RS-2021-II211817) and supported by the IITP-ICT Creative Consilience Program grant funded by the Korea government (MSIT) (IITP-2026-RS-2020-II201819).
\end{acks}

\clearpage
\appendix
\section*{Ethical Considerations}
Our research introduces \sys, a Rust-specific directed greybox fuzzer designed to improve the security of Rust programs by focusing fuzzing on areas where memory errors are most likely to occur.
Throughout the entire process—from design and implementation to evaluation and result disclosure (\eg bug report)—we carefully reviewed potential ethical issues in line with the ACM CCS guidelines and the Menlo Report, and found no ethical issues to report.

\textbf{Stakeholders.}
The potential stakeholders of our research include: (1) developers aiming to implement secure Rust programs, (2) end-users of Rust applications who benefit from improved security, (3) open-source project maintainers who may receive vulnerability reports, (4) attackers who could exploit vulnerabilities uncovered by this tool, and (5) security researchers focusing on vulnerability finding.

\textbf{Impacts (Ethical Principles).}
\emph{Beneficence:} our work focuses on efficiently testing memory-bug-prone areas in Rust programs, with the goal of improving overall software safety for developers and end-users. In addition, open-source project maintainers can benefit from receiving vulnerability reports derived from our research to improve the security of their projects.
\emph{Respect for Persons:} 
in this research, human participants and personal data processing were not involved. Additionally, all evaluations were performed in a locally managed, isolated environment to ensure no accidental contact with external systems or users.
\emph{Justice:} by openly sharing RustGo, we aim to provide equal opportunities for every user (not specific user and group) to strengthen the security of Rust applications.
\emph{Respect for Law and Public Interest:} 
all applications used in this research were open-source and fully compliant with their licenses; no license, terms-of-service, or legal violations occurred. Furthermore, each discovered vulnerability by RustGO was reported following the responsible disclosure process to prevent abuse by attackers.

\textbf{Impacts (Harms) and Mitigations.}
The potential harms of this study include: (1) the risk that discovered vulnerabilities could be exploited by attackers, (2) the burden placed on open-source project maintainers when vulnerabilities are reported, (3) the possibility that fuzzing experiments might affect external services, and (4) the risk that research artifacts could be misused for malicious purposes.
To address these potential harms, we followed responsible disclosure procedures when reporting vulnerabilities, providing maintainers with sufficient technical context to support efficient handling. All evaluations were conducted in a local, isolated environment to prevent unintended interaction with external services or users. Lastly, we ensured that all our released codes and data do not contain any exploit code that could enable direct attacks; instead, we share only findings intended for research and defensive use.

\textbf{Decision.}
We assessed the balance between ethical harms and benefits and determined that the benefits of this research clearly outweigh the risks. In terms of Beneficence, this research provides considerable security improvements by efficiently identifying memory vulnerabilities in Rust applications, benefiting both developers and end-users. At the same time, potential harm was minimized by conducting experiments in a local isolated environment, following responsible disclosure, and omitting exploit-ready code from release. In addition, this research involves neither human participants nor the collection of personal data, so no individual rights were violated. After considering all of these aspects, we conclude that the conduct and publication of this research yield greater benefits than harms.

\clearpage

\bibliographystyle{ACM-Reference-Format}
\bibliography{references}

\clearpage
\appendix
\section{Appendix}
\label{s:appendix}

\subsection{Open Science}
We provide all artifacts, including source code and datasets, in accordance with the ACM CCS open science policy to evaluate the contributions of this paper.

\textbf{Artifact Access.}
The artifacts of \sys can be accessed through the following anonymous repo:~\url{https://zenodo.org/records/21784601}.

\textbf{Artifact Contents and Reproduction.}
This artifact includes all the necessary code, scripts, and data to reproduce the evaluation and validate the claims of this paper. The provided artifacts include: (1) \sys source code: static analysis components, \sys fuzzer, and build scripts,
(2) Benchmark: 13 Rust programs with corresponding build scripts,
(3) Ported DGF: AFL, afl.rs, WindRanger, AFLGo, FishFuzz, AFLRUN, Lyso and PanicKiller source code and build scripts
(4) Scripts: scripts for static analysis validation, performance testing, and ablation study, and (5) README: instructions for building \sys and reproducing all evaluations.

\subsection{Evaluation Tables of \sys}
\label{apen:evaluation-result-details}

\begin{table}[h]
  \centering
  \caption{Dynamic Toggling Evaluation Result}
  \label{tbl:dynamic-toggling-evaluation}
  \renewcommand{\arraystretch}{0.9}
\resizebox{\columnwidth}{!}{%
\begin{tabular}{
>{\centering\arraybackslash}m{4.0cm}| 
>{\centering\arraybackslash}m{2.4cm}|
>{\centering\arraybackslash}m{2.4cm}
}
\toprule
\multirow{2}{*}{\textbf{Program}} 
& \multicolumn{1}{c|}{\textbf{AFLRUN}} 
& \multicolumn{1}{c}{\textbf{RustGo}} \\
\cline{2-3}
& \textbf{StdDev} 
& \textbf{StdDev}   \\

\midrule\midrule
\textbf{nom} 
& 1.57
& 1.52   \\

\midrule
\textbf{json} 
& 3.48 
& 1.73  \\

\midrule
\textbf{serde-yaml} 
& 2.89  
& 2.45  \\

\midrule
\textbf{unicode-process} 
& 3.52   
& 2.99  \\

\midrule
\textbf{unicode-streaming} 
& 2.02   
& 2.58 \\

\midrule
\textbf{data-encoding} 
& 2.49  
& 0.25  \\

\midrule
\textbf{chrono} 
& 14.95 
& 0.15  \\

\midrule
\textbf{mqtt-broker} 
& 2.99   
& 0.10  \\

\midrule
\textbf{qcms} 
& 2.21
& 1.88    \\

\midrule
\textbf{httparse}
& 4.45      
& 1.77   \\

\midrule
\textbf{quick-xml}
& 4.65
& 2.53    \\

\midrule
\textbf{lz4\_flex}
& 7.12  
& 1.56 \\

\midrule
\textbf{gimli}
& 11.10    
& 1.77    \\

\midrule\midrule
\textbf{Average} 
& 4.88
& 1.64 \\
\bottomrule
\end{tabular}%
}

\end{table}

\renewcommand{\arraystretch}{1.0} 
\begin{table}[t]
  \centering
  \caption{Results of inlining optimization on standard library and drop function calls. Drop Instances denote the number of drop instances in the benchmark programs. Std Calls and Drop Calls denote the number of standard library and drop function calls, respectively. Rate represents the percentage of function calls remaining after inlining optimization.}
  \label{tbl:drop-backup-data}
  \renewcommand{\arraystretch}{1.2}
\huge
\resizebox{\columnwidth}{!}{%
\begin{tabular}{
>{\centering\arraybackslash}m{4.5cm}| 
>{\centering\arraybackslash}m{2.0cm} |
>{\centering\arraybackslash}m{2.2cm} |
>{\centering\arraybackslash}m{2.2cm} | 
>{\centering\arraybackslash}m{2.2cm} |
>{\centering\arraybackslash}m{2.2cm} |
>{\centering\arraybackslash}m{2.2cm} |
>{\centering\arraybackslash}m{2.2cm} 
}
\toprule
\multirow[c]{2}{*}{\makecell[c]{\\\textbf{Program}}}
& \multirow[c]{2}{*}{\makecell[c]{\\\textbf{Drop}\\\textbf{Instance}}}
& \multicolumn{2}{c|}{\textbf{Before Inlining Opt}} 
& \multicolumn{4}{c}{\textbf{After Inlining Opt}} \\
\cline{3-8}
&
& \textbf{\makecell[c]{Std \\Calls (\#)}} & \textbf{\makecell[c]{Drop \\Calls (\#)}} 
& \textbf{\makecell[c]{Std \\Calls (\#)}} & \textbf{\makecell[c]{Rate \\ (\%)}} & \textbf{Drop Call (\#)} & \textbf{\makecell[c]{Rate \\ (\%)}}  \\

\midrule\midrule
\textbf{nom} 
& 436
& 4,284 & 763
& 4,154 & 96.97 & 762 & 99.87 \\

\midrule           
\textbf{json}   
& 468	
& 5,133	& 963	
& 4,998	& 97.37	& 960  & 99.69 \\

\midrule  
\textbf{serde-yaml}    
& 687	
& 6,543	& 1,513	
& 6,402	& 97.85	& 1,504	& 99.41 \\

\midrule  
\textbf{unicode-process}  
& 466	
& 4,406	& 814	
& 4,273	& 96.98	& 813	& 99.88 \\

\midrule  
\textbf{unicode-streaming}  
& 478	
& 4,366	& 815	
& 4,233	& 96.95	& 814	& 99.88 \\

\midrule  
\textbf{data-encoding}
& 450	
& 7,478	& 779	
& 7,346	& 98.23	& 778	& 99.87 \\

\midrule        
\textbf{chrono} 
& 461	
& 5,163	& 745	
& 5,089	& 98.57	& 744	& 99.87 \\

\midrule              
\textbf{mqtt-broker}
& 863	
& 6,237	& 1,653	
& 6,123	& 98.17	& 1,652	& 99.94 \\

\midrule      
\textbf{qcms} 
& 514	
& 6,045	& 1,226	
& 5,899	& 97.58	& 1,215	& 99.10 \\

\midrule                
\textbf{httparse}  
& 433	
& 4,182	& 725	
& 4,051	& 96.87	& 724	& 99.86 \\

\midrule           
\textbf{quick-xml}   
& 496	
& 4,876	& 869	
& 4,337	& 88.95	& 863	& 99.31 \\

\midrule    
\textbf{lz4\_flex}    
& 522	
& 4,106	& 886	
& 4,043	& 98.47	& 885	& 99.89 \\

\midrule         
\textbf{gimli}     
& 749	
& 5,816	& 1,205	
& 5,663	& 97.37	& 1,195	& 99.17 \\

\midrule\midrule
\textbf{Average} 
& 540	
& 5,280	& 997	
& 5,124	& 96.95	& 993	& 99.68 \\
\bottomrule
\end{tabular}%
}

\end{table}

\subsection{\cc{httparse} Case Study}
\label{apen:httparse-case-study}

We also analyze specific benchmarks to illustrate the individual impact of each component. 
In the \cc{httparse} benchmark, the number of target locations is reduced from 15 to 4 through semantic consolidation, improving 73.33\% target optimization rate and 269\% speedup (from $\times$ 6.81 of \rustgobase to $\times$ 4.12 of \rustgoopt), demonstrating the effectiveness of target optimization (\Sref{ss:DesignModule1}) in reducing redundant fuzzing efforts.

On the same application, \rustgobase removes only 49.94\% of unnecessary paths due to the over-approximation caused by Rust's standard library complexity, with standard library blocks accounting for 61.95\% (19,218 out of 31,020) of the total basic blocks.
In contrast, \rustgoprun achieves a path pruning rate of up to 77.83\% (24,143 out of 31,020 basic blocks pruned), reducing unnecessary fuzzing overhead up to $\times$ 2.40 (from $\times$ 4.12 of \rustgoopt to $\times$ 1.72 of \rustgoprun). These results validate that target-specific path pruning (\Sref{ss:DesignModule2}) improves fuzzing efficiency by eliminating unreachable paths.
The \cc{httparse} benchmark, in particular, demonstrates the most significant impact of the dynamic fair toggling fuzzing (\Sref{ss:DesignModule3}), improving 72.03\% speedup (from $\times$ 1.72 of \rustgoprun to $\times$ 1.00 of \sys). This is because \cc{httparse} retains four independent target locations even after post-dominant analysis, and dynamic pruning effectively distributes fuzzing resources across the corresponding distinct paths.

\subsection{Details of Comparison with Fuzz Driver Generator}
\label{apen:comparison-with-fuzz-driver-generator}

For RPG, it fails to detect 10 out of the 13 newly discovered vulnerabilities (a failure rate of 76.92\%). This failures stems from its reliance on a restricted argument type pool that primarily supports values directly constructible from fuzzing inputs, such as primitive types (\eg \cc{i32}). As a result, RPG cannot handle complex argument types that require prior initialization of internal fields (\eg \cc{key/value} pairs in \cc{BTreeMap}), and excludes the vulnerable functions from driver generation.
Moreover, RPG constructs API call sequences by selecting unsafe APIs and their related APIs for which argument types can be directly provided. As a result, it fails to capture the object and pointer flows required to trigger memory bugs (\eg freed objects accessed via raw pointers), leading in high false negatives, as the generated drivers do not capture the context required to trigger memory bugs.
%

For RUG, it fails to detect 9 out of the 12 newly discovered vulnerabilities (a failure rate of 75\%), excluding one program (dbn) where preprocess analysis is not applicable. Although RUG employs type dependency analysis to expand supported argument types (\eg trait), it still inherits a fundamental limitation shared with RPG. RUG can only generate inputs for arguments whose values can be assigned based solely on type information. As a result, RUG fails to construct inputs for complex argument types that require satisfying internal fields, and consequently excludes vulnerable functions requiring such arguments from driver generation during the preprocess stage.

For deepSURF, it fails to detect any of the 13 newly discovered memory violations (a failure rate of 100\%). It identifies unsafe code reachable from public safe APIs through unsafe reachability analysis and constructs API sequences by treating such these APIs as fuzzing entry points. However, since seven vulnerabilities are triggered through unsafe APIs that are reachable only via private or helper functions, deepSURF fails to construct sequences that include the vulnerability-triggering execution paths.

In the remaining cases, although deepSURF successfully identifies API sequences connected to the vulnerable unsafe code, it fails during the harness generation phase due to input construction limitations. Specifically, deepSURF generates inputs only for argument types that can be directly mapped from fuzzer provided input. Consequently, it cannot construct valid inputs for complex argument types that require additional initialization steps for internal fields (\eg \cc{Zipfile}). In contrast, \sys can handle these complex types by obtaining type information at runtime.

\section{Artifact Appendix}


\subsection{Abstract}

This artifact provides the implementation and evaluation environment of \sys, including eight comparison fuzzers (\ie AFL, AFL++, AFLGo, WindRanger, FishFuzz, AFLRUN, Lyso, and PanicKiller), five ablation variants (\ie \textsc{Dissafe}, \textsc{RG-base}, \textsc{RG-opt}, \textsc{RG-prun}, and \textsc{RG-sync}), 13 Rust benchmarks, and 26 deepSURF harnesses. It is distributed under the AGPL-3.0-or-later license and received the Artifacts Available, Artifacts Evaluated--Functional, and Results Reproduced badges at ACM CCS 2026.


\subsection{Description \& Requirements}

The source, build scripts, and evaluation materials are provided under \path{src/}, \path{build/}, and \path{evaluation/}, respectively. \path{README.md} describes the native workflow, while \path{GUIDE.md} describes the prebuilt Docker environment. We strongly recommend following the instructions in \path{README.md} and \path{GUIDE.md} when setting up and evaluating the artifact.



\subsubsection{Security, privacy, and ethical concerns}

There are no known security, privacy, or ethical concerns associated
with running this artifact. Fuzzing may trigger crashes or timeouts in
the included benchmarks, but these effects are confined to the benchmark
processes and Docker environment.


\subsubsection{How to access}

The latest version of the artifact is permanently available at
\url{https://zenodo.org/records/21784601}.


\subsubsection{Hardware dependencies}

No specialized hardware is required. The experiments reported in the paper were conducted on a machine equipped with a 48-core Intel(R) Xeon(R) Platinum 8268 CPU at 2.90\,GHz and 565\,GB of DDR4 memory.


\subsubsection{Software dependencies}
The artifact targets Linux/amd64 and was evaluated on Ubuntu 22.04.5 with Linux 6.8.0. It uses Rust 1.64.0, LLVM 14.0.6, SVF 2.4, and AFL++ 4.00c. The prebuilt Docker image containing the \sys components is also provided.


\subsubsection{Benchmarks}

The benchmark suite contains 13 Rust programs and is used for the main effectiveness and ablation experiments. The artifact additionally contains 26 real-world benchmark harnesses derived from deepSURF.


\subsection{Set Up}


\subsubsection{Installation}

The recommended setup uses the provided prebuilt Docker image. The main \sys components are already built under \path{/artifact/build/}. Detailed instructions are provided in \path{GUIDE.md}.

\begin{enumerate}

\item Load the Docker image:

\begin{Verbatim}
zstd -d 66-ccs26-artifacts-docker.tar.zst
docker load -i 66-ccs26-artifacts-docker.tar
\end{Verbatim}

\item Start the prebuilt \sys container:

\begin{Verbatim}
docker run -it 66-ccs26-artifacts:latest bash
\end{Verbatim}
\end{enumerate}

\noindent 
The Docker image allows evaluators to skip the time-consuming build of the main \sys toolchain. For the evaluation steps, follow the instructions in
\path{README.md}.

\subsubsection{Basic test}

To verify the prebuilt \sys environment, execute the following commands inside the container:

\begin{Verbatim}
cd /artifact
./smoke-test.sh
\end{Verbatim}

\noindent 
The smoke test checks the customized Rust compiler, static analysis, and \sys fuzzing engine, and builds a benchmark program (\ie \texttt{unicode-process}). Successful completion without errors indicates that the \sys toolchain and benchmark build pipeline are functioning correctly.


\subsection{Evaluation Workflow}


\subsubsection{Major claims}

\begin{itemize}

\item[(C1):]
The artifact provides the complete \sys evaluation environment, including eight comparison fuzzers and five ablation variants. This is validated by E1 and reported in Section~5.1 and Tables~1--2 of the paper.

\item[(C2):]
\sys eliminates 84.13\% of redundant targets and prunes an average of 78.49\% of irrelevant basic blocks. This is validated by E2 and reported in Sections~5.2--5.3 and Tables~3--4 of the paper.

\item[(C3):]
On the 13 standard Rust benchmarks, \sys is $\times$ 2.09--$\times$ 5.08 faster than the baseline fuzzers and $\times$ 1.36--$\times$ 3.27 faster than the five ablation variants. This is validated by E3 and reported in Sections~5.5--5.6 and Tables~5--6 of the paper.

\item[(C4):]
On the 26 deepSURF benchmarks, \sys is $\times$ 2.51--$\times$ 6.56 faster than the baseline fuzzers and $\times$ 2.02--$\times$ 3.87 faster than the five ablation variants. This is validated by E4 and reported in Section~5.5--5.6 and Table~7 of the paper.

\end{itemize}


\subsubsection{Experiments}
\begin{itemize}
\item[(E1):]
\textbf{Evaluation environment:} This experiment validates C1.

\textbf{Preparation:}
Initialize the \sys and evaluation environments. All commands below use \path{$EVAL=/artifact/evaluation/} as the evaluation root:
\begin{Verbatim}
source /artifact/env/env.sh
source $EVAL/evaluation-artifact/env.sh
\end{Verbatim}

\textbf{Execution:}
Initialize the comparison fuzzers and ablation variants:
\begin{Verbatim}
cd $EVAL/compared-fuzzer/build
./dgf-init.sh
cd $EVAL/ablation-variants
./variants-init.sh
\end{Verbatim}

\textbf{Results:}
Successful completion confirms that the comparison fuzzers and ablation
variants required for the evaluation are available and executable.


\item[(E2):]
\textbf{Static analysis evaluation:} This experiment validates C2.

\textbf{Preparation:}
Initialize the performance benchmarks:
\begin{Verbatim}
cd $EVAL/evaluation-artifact/performance
./performance_init.sh
\end{Verbatim}

\textbf{Execution:}
\begin{Verbatim}[breaklines=true,breakanywhere=true]
cd $EVAL/evaluation-artifact
cd performance/rustgo
./summarize_results.sh > rustgo_result.csv
\end{Verbatim}

\textbf{Results:}
The generated \path{rustgo_result.csv} summarizes the target optimization and path pruning results.


\item[(E3):]
\textbf{Benchmark evaluation:}
This experiment validates C3.

\textbf{Preparation:}
Use the benchmarks initialized in E2.

\textbf{Execution:}
\begin{Verbatim}
cd $EVAL/evaluation-artifact/performance
./run-performance-benchmark.sh
cd $EVAL/evaluation-artifact/ablation
./run-ablation-benchmark.sh
\end{Verbatim}

\textbf{Results:}
The wrappers summarize the target-reaching times recorded in
\path{reaching-log.txt} for the comparison fuzzers and ablation variants.

\item[(E4):]
\textbf{deepSURF benchmark evaluation:}
This experiment validates C4.

\textbf{Preparation:}
Initialize the deepSURF benchmarks:
\begin{Verbatim}
cd $EVAL/evaluation-artifact/deepsurf-benchmark
./deepsurf-benchmark-init.sh
\end{Verbatim}

\textbf{Execution:}

\begin{Verbatim}
cd $EVAL/evaluation-artifact/deepsurf-benchmark
cd performance 
./run-deepsurf-benchmark.sh
cd ../ablation 
./run-deepsurf-ablation.sh
\end{Verbatim}

\textbf{Results:}
The wrappers summarize the target-reaching times for the 26 deepSURF
benchmarks across the baseline fuzzers and ablation variants.

\end{itemize}

\subsection{Version}

Based on the LaTeX template for Artifact Evaluation V20220926.

\clearpage
\onecolumn
\section{Evaluation of \sys{} Figures}
\label{apen:evaluation-figures}

\begin{figure*}[!h]
  \centering
  \includegraphics[
    width=\textwidth,
    height=.30\textheight,
    keepaspectratio
  ]{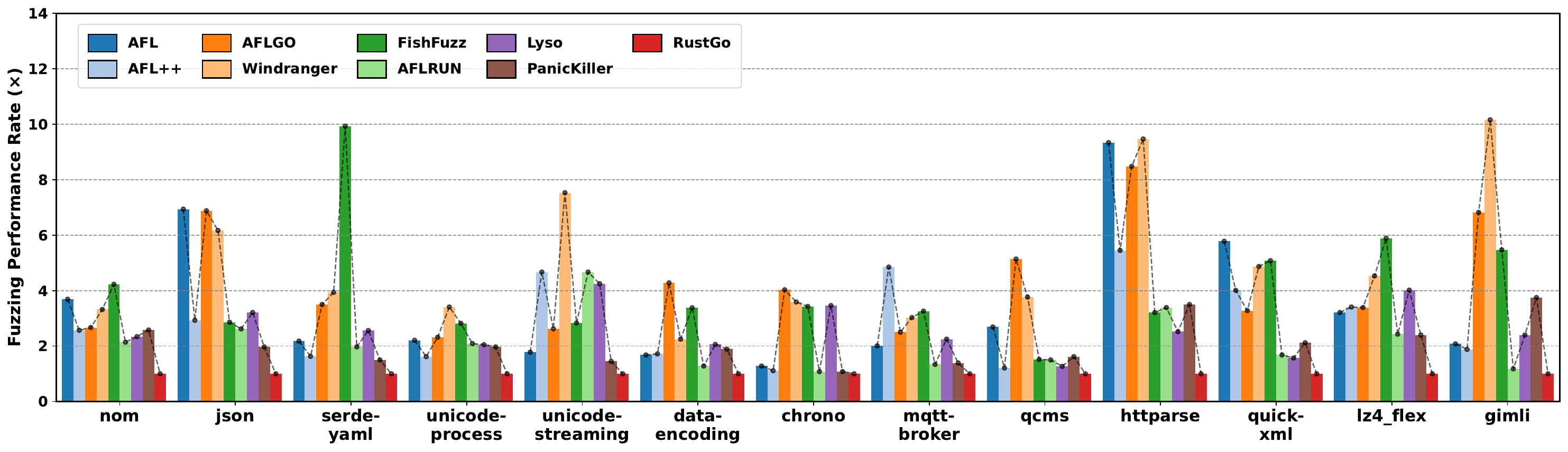}

  \Description{Grouped bar chart comparing the fuzzing performance
  rate of the evaluated fuzzers across the benchmark programs.}

  \caption{Comparison of exposure rate per application.}
  \label{fig:performance}
\end{figure*}

\begin{figure*}[!h]
  \centering
  \includegraphics[
    width=\textwidth,
    height=.30\textheight,
    keepaspectratio
  ]{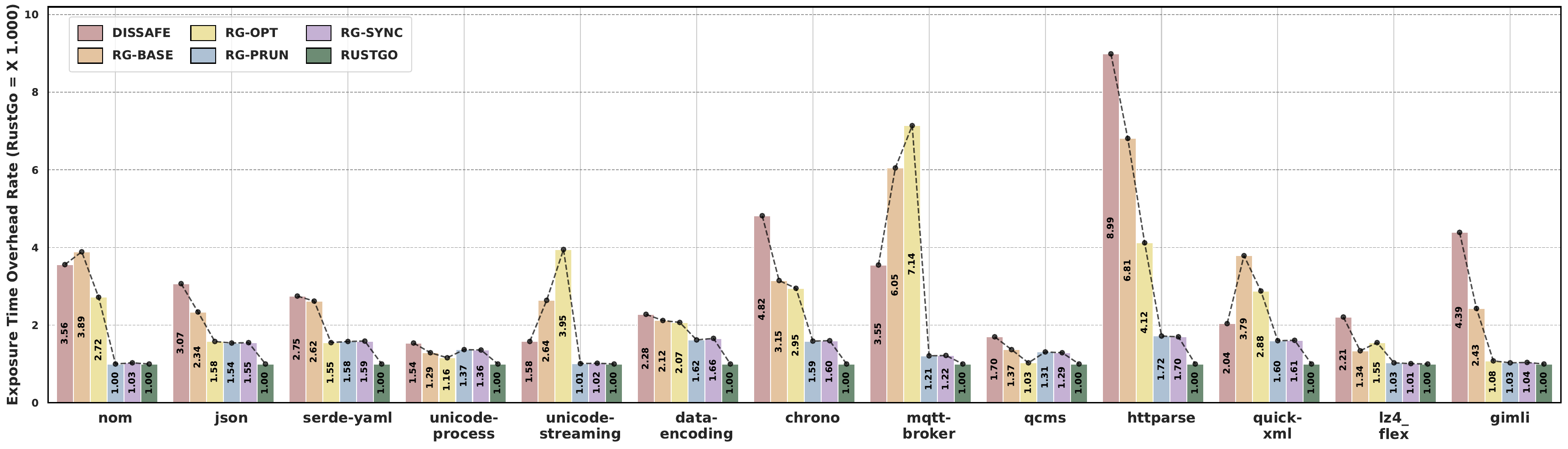}

  \Description{Grouped bar chart comparing the fuzzing performance
  of the RustGo variants across the benchmark programs.}

  \caption{Fuzzing performance of each variant.}
  \label{fig:ablation}
\end{figure*}

\end{document}